\documentclass[12pt]{article}
\usepackage{graphicx}
\usepackage{enumerate}
\usepackage{natbib}
\usepackage{url} 
\usepackage{epsfig}
\usepackage{amssymb, amsmath, amsthm}
\usepackage{hyperref}
\usepackage{xspace}

\usepackage[usenames,dvipsnames,svgnames,table]{xcolor}
\usepackage{subcaption}
\usepackage{algorithm}
\usepackage{algorithmic}
\floatname{algorithm}{Algorithm}

\usepackage{booktabs}  
\usepackage{pdflscape} 

\usepackage{longtable}
\usepackage{hhline}
\usepackage{float}

\DeclareMathOperator*{\argmin}{argmin}

\newcommand{\simiid}{\overset{iid}{\sim}}

\newcommand{\calC}{\mathcal{C}}

\newcommand{\calH}{\mathcal{H}}

\newcommand{\calP}{\mathcal{P}}

\newcommand{\calV}{\mathcal{V}}

\newcommand{\calY}{\mathcal{Y}}

\newcommand{\calMnr}{\mathcal{M}_{n,r}}
\newcommand{\calMnro}{\mathcal{M}^{\circ}_{n,r}} 

\newcommand{\bbP}{\mathbb{P}}
 
\newcommand{\bbR}{\mathbb{R}}
\newcommand{\bbS}{\mathbb{S}}
\newcommand{\bbE}{\mathbb{E}}
\newcommand{\bbN}{\mathbb{N}}

\newcommand{\bfA}{\boldsymbol{A}}
\newcommand{\bfB}{\boldsymbol{B}}

\newcommand{\bfH}{\boldsymbol{H}}
\newcommand{\bfI}{\boldsymbol{I}}
\newcommand{\bfQ}{\boldsymbol{Q}}
\newcommand{\bfR}{\boldsymbol{R}}
\newcommand{\bfS}{\boldsymbol{S}}

\newcommand{\bfU}{\boldsymbol{U}}
\newcommand{\bfV}{\boldsymbol{V}}
\newcommand{\bfW}{\boldsymbol{W}}
\newcommand{\bfX}{\boldsymbol{X}}
\newcommand{\bfY}{\boldsymbol{Y}}
\newcommand{\bfZ}{\boldsymbol{Z}}
\newcommand{\bfLambda}{\boldsymbol{\Lambda}}

\newcommand{\bfSigma}{\boldsymbol{\Sigma}}
\newcommand{\bfOmega}{\boldsymbol{\Omega}}

\newcommand{\bfx}{\boldsymbol{x}}
\newcommand{\bfy}{\boldsymbol{y}}

\newcommand{\bfone}{\boldsymbol{1}}
\newcommand{\bfzero}{\boldsymbol{0}}
\newcommand{\bft}{\boldsymbol{t}}
\newcommand{\bfxi}{\boldsymbol{\xi}}

\newcommand{\Frechet}{Fr\'{e}chet\xspace}

\newcommand{\Log}{\mathrm{Log}}

\newcommand{\BhatF}{\widehat{\bfB}_{\mathrm{F}}}
\newcommand{\BF}{\bfB_{\mathrm{F}}}
\newcommand{\VarFhat}{\widehat{\mathrm{Var}}_{F}}

\definecolor{wass-mean}{HTML}{CC79A7}
\definecolor{wass-meds}{HTML}{0072B2}
\definecolor{wass-class1}{HTML}{56B4E9}
\definecolor{wass-class2}{HTML}{D55E00}

\newtheorem{theorem}{Theorem}
\newtheorem{lemma}[theorem]{Lemma}

\newtheorem{corollary}[theorem]{Corollary}

\newcommand{\blind}{1}

\begin{document}

\def\spacingset#1{\renewcommand{\baselinestretch}%
{#1}\small\normalsize} \spacingset{1}


\if1\blind
{
  \title{\bf Quotient-Based Posterior Analysis for Euclidean Latent Space Models}
  \author{Kisung You\\
    Department of Mathematics, Baruch College, City University of New York\\
    and \\
    Mauro Giuffr\`{e}\\
    Department of Biomedical Informatics \& Data Science, Yale University\\
    Department of Medicine, Surgery, and Health Sciences, University of Trieste\\
    }
  \maketitle
} \fi

\if0\blind
{
  \bigskip
  \bigskip
  \bigskip
  \begin{center}
    {\LARGE\bf Quotient-Based Posterior Analysis for Euclidean Latent Space Models}
\end{center}
  \medskip
} \fi

\bigskip
\begin{abstract}
Latent space models are widely used in statistical network analysis and are often fit by Markov
chain Monte Carlo. However, posterior summaries of latent coordinates are not canonical because the
likelihood depends only on pairwise distances and is invariant under rigid motions of the latent
space. Standard post hoc alignment can aid visualization, but the resulting summaries depend on an
arbitrary reference configuration.  We propose a quotient-based posterior analysis for Euclidean latent space models using the centered
Gram map, which represents identifiable latent structure while removing nonidentifiability. This
yields intrinsic posterior summaries of mean structure and uncertainty that can be computed directly
from posterior samples, together with basic theoretical guarantees including canonicality, existence,
and stability.  Through simulations and analyses of the Florentine marriage network and a statisticians'
coauthorship network, the proposed framework clarifies when alignment-based summaries are stable,
when they become reference-sensitive, and which nodes or relationships are weakly identified. These
results show how coherent posterior analysis can reveal latent relational structure beyond a single
embedding.
\end{abstract}

\noindent%

\newpage
\spacingset{1.9} 

\section{Introduction}\label{sec:intro}

Latent space models (LSMs) provide a flexible and interpretable framework for statistical network
analysis \citep{hoff_2002_LatentSpaceApproaches}. Let $G=(V,E)$ be an undirected graph with node set
$V=\{1,\ldots,n\}$ and observed adjacency matrix
\[
\bfA=(A_{ij})\in\{0,1\}^{n\times n},\qquad A_{ij}=A_{ji},\ \ A_{ii}=0,
\]
where $A_{ij}=1$ indicates an edge between nodes $i$ and $j$. In the standard Euclidean LSM, each
node $i$ has an unobserved latent position $\bfx_i\in\bbR^r$, collected as
$ \bfX=(\bfx_1,\ldots,\bfx_n)^\top\in\bbR^{n\times r}.$
Conditional on $\bfX$, edges are modeled as conditionally independent with
\begin{equation}\label{eq:lsm}
\bbP(A_{ij}=1 \mid \bfX,\alpha)
= g\!\left(\alpha - \|\bfx_i - \bfx_j\|\right),
\qquad 1\le i<j\le n,
\end{equation}
where $g(\cdot)$ is an inverse link function and $\alpha$ is an intercept. Since
their introduction by \citet{hoff_2002_LatentSpaceApproaches}, Euclidean LSMs have been studied
largely in Bayesian form, with priors on $(\bfX,\alpha)$ and posterior inference based on Markov
chain Monte Carlo (MCMC). A typical specification is
$\bfx_i \simiid \mathcal{N}(\bfzero,\sigma^2 \bfI_r)$ and
$\alpha \sim \mathcal{N}(0,\tau^2)$, although our development does not depend on a particular prior.
Throughout, we use LSM to mean Euclidean LSM.

In practice, posterior draws are only intermediate objects. Scientific conclusions are conveyed
through posterior means, credible regions, uncertainty displays, and predictive summaries. Such
summaries are meaningful only if they target identifiable structure. In LSMs, however, the
likelihood depends on $\bfX$ only through pairwise distances and is therefore invariant under rigid
motions of the latent space. Under common isotropic priors, the posterior typically retains
rotational and reflectional invariance. Consequently, coordinate-dependent summaries such as
posterior means of $\bfx_i$ or coordinate-wise credible intervals are not canonical as they may change
under arbitrary reorientation of the latent space even when the posterior is tightly concentrated in
relational terms. 

A common practical response is to center posterior draws and align them to a reference
configuration via orthogonal Procrustes matching
\citep{hoff_2002_LatentSpaceApproaches, gower_2004_ProcrustesProblems}. This is useful for
visualization, but it is unsatisfactory as an inferential primitive because the resulting summaries
depend on an external reference, such as the first draw or a point estimate, that is not part of the
Bayesian model.

We address this problem by carrying out posterior analysis on a representation that encodes only
identifiable latent structure. Specifically, we map each latent configuration $\bfX$ to its centered
Gram matrix
$
\Phi(\bfX):=\bfH\bfX\bfX^\top\bfH,
$
where $\bfH$ is the centering matrix, where the map is invariant under rigid motions and preserves all
pairwise distances. Pushing the posterior through $\Phi$ yields an induced posterior distribution on
a quotient state space of centered, low-rank positive semidefinite matrices. Posterior summaries are
then defined directly on this quotient space, and hence depend only on the induced posterior
distribution of identifiable structure, not an arbitrarily specified reference outside of the model.

A natural question is why translation is removed as well, given that common Gaussian priors break
translation invariance of the posterior. The point is that our inferential target is not the
absolute origin of the latent coordinate system, which is prior-dependent, but the relational latent
structure encoded by pairwise distances. Centering chooses a canonical representative of each
distance-equivalent configuration while preserving all pairwise distances exactly. In this sense,
quotienting out translation is a choice of estimand rather than a claim that translation remains a
posterior symmetry.

This quotient perspective naturally prioritizes rigid-motion-invariant inferential quantities and
builds on classical distance geometry and multidimensional scaling, where Gram matrices provide
rigid-motion invariants of point configurations
\citep{gower_1969_MinimumSpanningTrees, borg_1997_ModernMultidimensionalScaling}. Geometrically,
it also aligns with quotient-manifold treatments of fixed-rank positive semidefinite matrices under
orthogonal group actions
\citep{absil_2008_OptimizationAlgorithmsMatrix, journee_2010_LowRankOptimizationCone, massart_2020_QuotientGeometrySimple},
while our use of geometric summaries draws on statistics on manifolds
\citep{bhattacharya_2012_NonparametricInferenceManifolds, patrangenaru_2016_NonparametricStatisticsManifolds, pennec_2020_RiemannianGeometricStatistics}. Conceptually, the issue is analogous to label switching in mixture models, where meaningful posterior summaries must be invariant to permutations rather than tied to arbitrary labels
\citep{celeux_2000_ComputationalInferentialDifficulties, stephens_2000_DealingLabelSwitching, jasra_2005_MarkovChainMonte}.

This paper makes four contributions.
\begin{enumerate}
\item We formalize a quotient state space for LSMs based on centered Gram matrices,
capturing identifiable latent structure while removing rigid-motion nonidentifiability.
\item We define intrinsic posterior summaries on this space, including a \Frechet posterior mean,
global dispersion, intrinsic credible regions, dyad-level summaries for distances and edge
probabilities, and node-level uncertainty indices.
\item We provide practical algorithms to compute these summaries directly from posterior draws, so
the framework can be used as a post-processing layer on top of existing MCMC routines for 
LSMs.
\item We establish basic theoretical properties, including canonicality, existence of \Frechet means
on a closed state space, stability under perturbations of the posterior law, and consistency of
distance-based inference under posterior concentration.
\end{enumerate}

The remainder of the paper is organized as follows. Section~\ref{sec:background} reviews the
identifiability issues underlying LSMs and clarifies why alignment-based summaries are not suitable 
inferential primitives. Section~\ref{sec:quotient} develops the centered Gram representation and the
quotient geometry used throughout. Section~\ref{sec:summaries} introduces intrinsic posterior
summaries and computational procedures, while Section~\ref{sec:theory} establishes their basic
theoretical properties. Section~\ref{sec:experiments} reports a simulation study together with
analyses of the Florentine marriage network and a statisticians' coauthorship network, illustrating
both the instability of reference-based summaries and the interpretability of quotient-based
uncertainty. Section~\ref{sec:conclusion} concludes with a discussion of implications and
extensions.


\section{Motivation and Background}\label{sec:background}

We begin with notation. We write $\|\cdot\|_F$ for the Frobenius norm and $\|\cdot\|$ for the $\ell_2$ norm. The orthogonal group is
$O(r)=\{\bfR\in\bbR^{r\times r}:\bfR\bfR^\top=\bfI_r\}$, where $\bfI_r$ is the $r\times r$
identity matrix. The centering matrix is
\[
\bfH = \bfI - \frac{1}{n}\bfone_n\bfone_n^\top,
\]
where $\bfone_n$ is the length-$n$ vector of ones, so that $\bfH\bfone_n=\bfzero$ and
$\bfH^\top=\bfH$. For $n\in\bbN$, we write $[n]:=\{1,\ldots,n\}$.

\subsection{Identifiability in latent space models}

In LSMs of the form \eqref{eq:lsm}, the likelihood depends on $\bfX$ only through pairwise distances.
Thus, for any orthogonal matrix $\bfR\in O(r)$ and translation vector $\bft\in\bbR^r$,
\[
\| (\bfR^\top \bfx_i + \bft) - (\bfR^\top x_j + \bft) \| = \|x_i - x_j\|,
\]
so the likelihood is invariant under the rigid-motion group action
\[
g\cdot \bfX := \bfX\bfR + \bfone_n \bft^\top,\qquad g=(\bfR,\bft)\in O(r)\times\bbR^r.
\]
This invariance is structural, directly rooted in using inter-point distances in the model.

In its Bayesian form, LSMs typically employ priors that concentrate latent positions near the origin, most often
independent isotropic Gaussians. Such priors break translation invariance of the posterior by penalizing configurations far from the origin. Consequently, translation is not a posterior symmetry even though it remains a likelihood symmetry. This distinction helps to clarify why centering posterior
draws is coherent and widely used, rather than an arbitrary post hoc convention.

By contrast, isotropic priors do not break rotational or reflectional invariance. Both likelihood and
prior remain invariant under the map $\bfX\mapsto \bfX\bfR$ for $\bfR\in O(r)$, so the posterior assigns equal probability
to all rotated and/or reflected versions of a configuration. This property makes the orientation as a fundamentally unidentifiable object.

\subsection{Posterior sampling versus posterior summarization}

Rotational non-identifiability does not pose a problem for MCMC sampling. The chain may explore
equivalent orientations, and rotational variability in the draws is rather a faithful representation of
posterior uncertainty under the model. Difficulties arise only when one attempts to summarize the posterior. Many common summaries are defined in the ambient Euclidean coordinate system, e.g., $\bbE[\bfx_i\vert \bfA]$ or coordinate-wise
credible intervals. However, the posterior distribution is not naturally a distribution on
$\bbR^{n\times r}$ with identifiable coordinates. Instead, it is a distribution on equivalence classes
under orthogonal transformations. As a consequence, coordinate-dependent summaries are not uniquely
determined by the posterior distribution alone.

Consider a simple thought experiment to illustrate this point. Suppose the posterior concentrates sharply near a
single configuration $\bfX^\star$ up to rotation, so that draws look like $\bfX^{(m)}\approx \bfX^\star \bfR_m$
with $\bfR_m$ varying roughly uniformly in $O(r)$. Then the coordinate-wise posterior mean would be approximately zero by symmetry, even though the latent structure is highly concentrated. This example asserts that the issue is not uncertainty. Indeed, it shows that even the simplest posterior summary measure such as the mean is taken in an inappropriate non-identifiable coordinate system, which leads to an unreliable conclusion.

\subsection{Limitations of alignment-based summaries}

As noted earlier, a widely used remedy is to align posterior draws using orthogonal Procrustes matching. After centering, each draw is rotated to match a chosen reference configuration, and posterior summaries
are computed in the aligned coordinates. While this approach is  effective for visualization, it has limitations as an inferential
primitive. First, the alignment depends on an arbitrary reference configuration, which is not part of the model. Second, the alignment transforms each draw in a draw-specific manner, and the resulting summary
is generally not a functional of the posterior distribution alone. Lastly, in multimodal posteriors on the quotient, e.g., when the identifiable structure itself is
uncertain, forced alignment can artificially reduce apparent variability or introduce artifacts by
collapsing distinct structures into a single coordinate frame. 

These observations do not imply that alignment is wrong for plotting. Rather, they imply that alignment should be a visualization step applied after inference is defined intrinsically, the latter of which is not attainable via post hoc alignment. Hence, the discussion above leads to the desirable principle that posterior summaries for LSMs should be defined on a state space that reflects the identifiability
structure of the model. Summaries should be functionals of the posterior distribution on this space,
and only then may one choose an arbitrary orientation for visualization. In the LSM, the identifiable information is contained in pairwise distances, and centered Gram
matrices provide a classical representation that preserves these distances while removing rigid-motion degrees of freedom.

\section{A Quotient Representation for Latent Space Models}\label{sec:quotient}

\subsection{Centered Gram map and invariance}\label{subsec:gram}

Let $\bfX\in\bbR^{n\times r}$ and define the centered configuration $\tilde \bfX := \bfH \bfX$.
The associated centered Gram matrix is
\begin{equation}\label{eq:gram}
\bfB := \Phi(\bfX) = \tilde \bfX\tilde \bfX^\top = \bfH \bfX \bfX^\top \bfH.
\end{equation}
This representation is classical in distance geometry and multidimensional scaling
\citep{gower_1969_MinimumSpanningTrees, borg_1997_ModernMultidimensionalScaling}. 
The centered Gram map has three basic properties: for any $\bft\in\bbR^r$ and $\bfR\in O(r)$,
\[
\Phi(\bfX+\bfone_n\bft^\top)=\Phi(\bfX),\qquad
\Phi(\bfX\bfR)=\Phi(\bfX),
\]
and, for all $i,j$,
\begin{equation}\label{eq:dist-from-B}
\|\bfx_i-\bfx_j\|^2
=
\|\tilde\bfx_i-\tilde\bfx_j\|^2
=
\bfB_{ii}+\bfB_{jj}-2\bfB_{ij}.
\end{equation}
Thus the map $\Phi$ removes nuisance symmetries while preserving  the likelihood-relevant geometry exactly.

\subsection{Connection to squared distance matrices and classical MDS}\label{subsec:mds}

Let $\Delta(\bfX)\in\bbR^{n\times n}$ denote the squared distance matrix with entries
$\Delta_{ij}(\bfX)=\|\bfx_i-\bfx_j\|^2$. A standard identity in classical MDS is
\begin{equation}\label{eq:B-from-Delta}
\bfB = -\frac{1}{2}\,\bfH\,\Delta(\bfX)\,\bfH.
\end{equation}
Hence the centered Gram matrix is the doubly centered version of the squared distance matrix.
Conversely, \eqref{eq:dist-from-B} recovers all squared distances from $\bfB$. In this sense,
$\bfB$ is the canonical Euclidean embedding operator associated with distances, which makes it a
natural object for inference when distances are identifiable. Equation \eqref{eq:B-from-Delta} is the standard classical MDS identity, whose derivation is
given in the Supplementary Material for completeness.

\subsection{Identifiability and the quotient state space}\label{subsec:ident-state}

The centered Gram map $\Phi(\bfX)=\bfH \bfX \bfX^\top \bfH$ removes the rigid-motion
non-identifiabilities of LSMs while preserving the pairwise geometry that enters the likelihood.
Once centered, the Gram matrix identifies the latent configuration up to the orthogonal action that
remains unidentifiable under isotropic priors. The next lemma formalizes this point.

\begin{lemma}[Gram characterization]\label{lem:gram}
Let $\bfX,\bfW\in\bbR^{n\times r}$ and write $\tilde \bfX=\bfH \bfX$, $\tilde \bfW=\bfH \bfW$.
Suppose $\mathrm{rank}(\tilde \bfX)=\mathrm{rank}(\tilde \bfW)=r$. Then
\[
\tilde \bfX\tilde \bfX^\top = \tilde \bfW\tilde \bfW^\top
\quad\Longleftrightarrow\quad
\tilde \bfX = \tilde \bfW \bfR\quad \text{for some } \bfR\in O(r).
\]
Equivalently,
\[
\Phi(\bfX)=\Phi(\bfW)
\quad\Longleftrightarrow\quad
\bfX = \bfW \bfR + \bfone_n \bft^\top\quad\text{for some } (\bfR,\bft)\in O(r)\times\bbR^r.
\]
\end{lemma}

Lemma~\ref{lem:gram} shows that $\Phi(\bfX)$ captures exactly the rigid-motion equivalence class of
$\bfX$ under the full-rank condition. This motivates viewing posterior inference for LSMs as
inference on the image of $\Phi$, rather than on arbitrary coordinate representatives.

We therefore define the inferential state space of identifiable latent structure by
\begin{equation}\label{eq:state-space}
\calMnr := \left\{
\bfB\in\bbR^{n\times n} \;:\; \bfB\succeq 0,\ \mathrm{rank}(\bfB)\le r,\ \bfB\bfone_n=\bfzero
\right\}.
\end{equation}
The rank constraint $\mathrm{rank}(\bfB)\le r$ makes $\calMnr$ closed, which is convenient for
existence results such as \Frechet means. The rank-$r$ stratum
\[
\calMnro := \{\bfB\in\calMnr:\ \mathrm{rank}(\bfB)=r\}
\]
is a smooth manifold and is the typical support of draws from an $r$-dimensional latent space model. Every $\bfB\in\calMnro$ admits a factorization $\bfB=\bfY\bfY^\top$ with
$\bfY\in\bbR^{n\times r}$, $\bfY^\top\bfone_n=\bfzero$, and $\mathrm{rank}(\bfY)=r$, unique up to
right multiplication by an orthogonal matrix: $\bfY\sim \bfY\bfR$ for $\bfR\in O(r)$. This quotient
structure is the geometric manifestation of rotational and reflectional non-identifiability.

For an LSM with latent dimension $r$, every draw automatically satisfies
$\mathrm{rank}(\Phi(\bfX^{(m)}))\le r$. Allowing rank deficiency in \eqref{eq:state-space}
therefore avoids boundary issues while remaining faithful to what the model can generate. When
needed, one may assume posterior mass concentrates on $\calMnro$ away from the rank-deficient
boundary.

\subsection{Quotient distance and computation}\label{subsec:distance}

We next equip the space of identifiable latent structures with a distance intrinsic to the quotient
representation. Every $\bfB\in\calMnr$ is centered and positive semidefinite with
$\mathrm{rank}(\bfB)\le r$, and hence admits a factorization
\[
\bfB=\bfY\bfY^\top,\qquad \bfY\in\bbR^{n\times r}.
\]
Moreover, the centering constraint $\bfB\bfone_n=\bfzero$ allows us to choose a centered factor
without loss of generality. Indeed, for any factor $\bfY$,
\[
(\bfH\bfY)(\bfH\bfY)^\top = \bfH\bfY\bfY^\top\bfH = \bfH\bfB\bfH = \bfB
\quad\text{and}\quad
(\bfH\bfY)^\top\bfone_n=\bfY^\top(\bfH\bfone_n)=\bfzero.
\]
Thus we represent $\bfB$ throughout using factors satisfying $\bfY^\top\bfone_n=\bfzero$. When
$\mathrm{rank}(\bfB)<r$, some columns of $\bfY$ may be redundant or zero.

Given $\bfB_1,\bfB_2\in\calMnr$ with 
$\bfB_k=\bfY_k\bfY_k^\top$, one defines the quotient distance
\begin{equation}\label{eq:quotient-distance}
d(\bfB_1,\bfB_2)
:=
\min_{\bfR\in O(r)} \|\bfY_1-\bfY_2\bfR\|_F.
\end{equation}
This depends only on $\bfB_1$ and $\bfB_2$ where the minimization over $O(r)$ removes the remaining
gauge freedom in the factorization. Indeed, the minimizer in \eqref{eq:quotient-distance} is obtained by a small orthogonal Procrustes problem
\citep{gower_2004_ProcrustesProblems}. Let $\bfY_2^\top\bfY_1=\bfU\bfSigma\bfV^\top$ be a singular
value decomposition. Then
\begin{equation}\label{eq:procrustes}
\bfR^\star=\bfU\bfV^\top,\qquad
d(\bfB_1,\bfB_2)=\|\bfY_1-\bfY_2\bfR^\star\|_F.
\end{equation}
If $\bfY_2^\top\bfY_1$ is rank deficient, $\bfR^\star$ need not be unique, but the minimum value is.
Equivalently,
\[
d^2(\bfB_1,\bfB_2)
=
\|\bfY_1\|_F^2+\|\bfY_2\|_F^2-2\,\mathrm{tr}(\bfSigma),
\]
so the distance depends only on the singular values of $\bfY_2^\top\bfY_1$.
Since $r$ is typically small, e.g., $r=2$ or $3$ for visualization, computing this distance is
inexpensive: forming $\bfY_2^\top\bfY_1$ costs $O(nr^2)$ and the $r\times r$ SVD costs $O(r^3)$.

\subsection{Geometric operations on the rank-$r$ stratum}\label{subsec:geomops}

The quotient distance \eqref{eq:quotient-distance} is defined on the closed space $\calMnr$,
including points with $\mathrm{rank}(\bfB)<r$. For the posterior summaries developed later, local
linearization is most naturally formulated on the smooth rank-$r$ stratum
\[
\calMnro := \{\bfB\in\calMnr:\ \mathrm{rank}(\bfB)=r\},
\]
where the quotient structure is smooth and factor representatives have full column rank. On $\calMnro$, we work with the total space
\[
\bar{\calY}_{n,r}
:=
\{\bfY\in\bbR^{n\times r}:\ \bfY^\top\bfone_n=\bfzero,\ \mathrm{rank}(\bfY)=r\},
\]
and identify $\bfB=\bfY\bfY^\top$ under the equivalence $\bfY\sim\bfY\bfR$ for $\bfR\in O(r)$.
Thus $\calMnro$ is the quotient of $\bar{\calY}_{n,r}$ under the right action of $O(r)$. The
full-rank condition ensures that $\bfS:=\bfY^\top\bfY$ is positive definite, which in turn makes
the Lyapunov equation used below well posed. A tangent vector at $\bfY$ is any
$\bfZ\in\bbR^{n\times r}$ satisfying the linearized centering constraint
$\bfZ^\top\bfone_n=\bfzero$.

The orthogonal gauge invariance induces the vertical space
\[
\calV_{\bfY}=\{\bfY\bfOmega:\ \bfOmega^\top=-\bfOmega\in\bbR^{r\times r}\},
\]
which consists of directions that do not change $\bfB=\bfY\bfY^\top$. Under the Frobenius inner
product, a standard horizontal space is
\[
\calH_{\bfY}
=
\{\bfZ:\ \bfZ^\top\bfone_n=\bfzero,\ \bfY^\top\bfZ \text{ is symmetric}\},
\]
whose elements represent identifiable variation in $\bfB$.

Given any tangent vector $\bfZ$ with $\bfZ^\top\bfone_n=\bfzero$, its horizontal projection can be
written as
\[
\mathrm{Proj}^{\mathrm{hor}}_{\bfY}(\bfZ)=\bfZ-\bfY\bfOmega,
\]
where $\bfOmega^\top=-\bfOmega$ solves the Lyapunov equation
\citep{simoncini_2016_ComputationalMethodsLinear}
\begin{equation}\label{eq:sylvester}
\bfS\bfOmega+\bfOmega\bfS = \bfY^\top\bfZ - \bfZ^\top\bfY,
\qquad \bfS:=\bfY^\top\bfY\in\bbS_{++}^r.
\end{equation}

When optimization on the manifold is needed, we use the retraction
\citep{absil_2008_OptimizationAlgorithmsMatrix}
\begin{equation}\label{eq:retraction}
\mathrm{Retr}_{\bfY}(\bfZ):=\bfH(\bfY+\bfZ),
\end{equation}
which enforces the centering constraint numerically and coincides with a straight-line update in the
factor space followed by recentering. Similarly, given $\bfY_1,\bfY_2\in\bar{\calY}_{n,r}$, let
$\bfR^\star$ be the Procrustes solution from \eqref{eq:procrustes} and define the log lift at
$\bfY_1$ by
\begin{equation}\label{eq:loglift}
\Log_{\bfY_1}(\bfY_2)
:=
\mathrm{Proj}^{\mathrm{hor}}_{\bfY_1}(\bfY_2\bfR^\star-\bfY_1)\in \calH_{\bfY_1}.
\end{equation}
This provides a symmetry-respecting local coordinate for comparing $\bfB_2$ to $\bfB_1$ along
identifiable directions. We note that  the aligned difference $\bfZ:=\bfY_2\bfR^\star-\bfY_1$ is already horizontal at $\bfY_1$,
since $\bfY_1^\top\bfY_2\bfR^\star$ is symmetric. Hence
\[
\Log_{\bfY_1}(\bfY_2)=\bfZ
\qquad\text{and}\qquad
d(\bfB_1,\bfB_2)=\|\bfY_1-\bfY_2\bfR^\star\|_F=\|\Log_{\bfY_1}(\bfY_2)\|_F.
\]
Thus the quotient distance \eqref{eq:quotient-distance} and the log lift
\eqref{eq:loglift} are compatible on the rank-$r$ stratum $\calMnro$. On the rank-deficient
boundary, the quotient is no longer smooth and log-type constructions require additional care.
\section{Posterior Summaries}\label{sec:summaries}

This section develops posterior summaries for LSMs that are coherent with the quotient
representation introduced in Section~\ref{sec:quotient}. Let
$\bfX^{(1)},\ldots,\bfX^{(M)}$ denote posterior draws of the latent configuration from
$\Pi(\,\cdot\,\vert \bfA)$, and let
$\alpha^{(1)},\ldots,\alpha^{(M)}$ denote the corresponding intercept draws when needed.
Each latent draw induces a centered Gram matrix
\[
\bfB^{(m)} := \Phi(\bfX^{(m)})=\bfH \bfX^{(m)}\bfX^{(m)\top}\bfH,\qquad m=1,\ldots,M,
\]
which is invariant under rigid motions and preserves all pairwise distances. Defining
\[
\bfY^{(m)}:=\bfH\bfX^{(m)}\in\bbR^{n\times r},
\qquad (\bfY^{(m)})^\top\bfone_n=\bfzero,
\]
we can equivalently write $\bfB^{(m)}=\bfY^{(m)}(\bfY^{(m)})^\top$. Throughout, we treat these
draws as given and focus on coherent post-processing of MCMC output. Thus the contribution here is a
posterior-analysis layer that can be combined with any MCMC routine for LSMs.

Two complementary modes of inference are useful. The first is intrinsic Monte Carlo inference,
which works directly with the empirical distribution of $\{\bfB^{(m)}\}_{m=1}^M$ or its functionals.
The second is local linearization on the rank-$r$ stratum $\calMnro$, which provides tractable
second-order approximations to posterior variability. Both are defined without using coordinate
alignment as an inferential primitive.

\subsection{Posterior mean}\label{subsec:mean}

A natural posterior point summary should represent the center of the induced posterior on
$(\calMnr,d)$. Since Euclidean means are characterized as minimizers of sums of squared distances,
we define the sample \Frechet mean of identifiable structure by
\begin{equation}\label{eq:frechet-mean}
\BhatF \in \argmin_{\bfB\in\calMnr}\ \frac{1}{M}\sum_{m=1}^M d^2(\bfB,\bfB^{(m)}).
\end{equation}
By construction, $\BhatF$ is invariant under rigid motions and depends only on the empirical
posterior distribution of $\bfB$.

For computation, it is convenient to work with factor representatives. Writing
$\bfB=\bfY\bfY^\top$ with $\bfY^\top\bfone_n=\bfzero$, the \Frechet objective becomes
\[
F(\bfY)
=
\frac{1}{M}\sum_{m=1}^M
\min_{\bfR_m\in O(r)} \|\bfY-\bfY^{(m)}\bfR_m\|_F^2,
\qquad \text{subject to }\bfY^\top\bfone_n=\bfzero.
\]
Although not jointly convex, $F$ is smooth on the rank-$r$ stratum in a quotient sense and can be
handled by first-order methods.

Starting from a current iterate $\bfY$, we solve for each $m\in[M]$ the orthogonal Procrustes
problem
\[
\bfR_m^\star \in \argmin_{\bfR\in O(r)}\|\bfY-\bfY^{(m)}\bfR\|_F,
\]
whose solution is obtained via an SVD of $(\bfY^{(m)})^\top\bfY$. We then compute
\[
\bar{\bfY} := \frac{1}{M}\sum_{m=1}^M \bfY^{(m)}\bfR_m^\star,\qquad
\bfZ := \bar{\bfY}-\bfY.
\]
Because the factorization is defined only up to right-orthogonal transformations, $\bfZ$ contains
nonidentifiable components. We remove these by horizontal projection,
$\bfZ_{\mathrm{hor}} := \mathrm{Proj}^{\mathrm{hor}}_{\bfY}(\bfZ)$, obtained from the Lyapunov
equation \eqref{eq:sylvester}, and update via the retraction \eqref{eq:retraction}
$
\bfY^+ = \mathrm{Retr}_{\bfY}(\eta\,\bfZ_{\mathrm{hor}}),
$
with step size $\eta>0$. After convergence, the posterior mean is reported as
$
\BhatF=\hat{\bfY}\hat{\bfY}^\top.
$ A full algorithmic description is provided in the Supplementary Material.

\subsection{Global and local posterior variability}\label{subsec:var}

After defining an intrinsic posterior mean, we quantify uncertainty in two ways: a global dispersion
measure and a local second-order approximation around $\BhatF$.

A global intrinsic dispersion measure is the sample \Frechet variation,
\begin{equation}\label{eq:frechet-var}
\VarFhat := \frac{1}{M}\sum_{m=1}^M d^2(\BhatF,\bfB^{(m)}),
\end{equation}
the attained minimum of the \Frechet objective. This summarizes overall posterior spread around
$\BhatF$ without reference to any coordinate system.

A natural global credible region is the intrinsic ball
\[
\calC_{1-\alpha} := \{\bfB\in\calMnr:\ d(\bfB,\BhatF)\le r_{1-\alpha}\},
\]
where $r_{1-\alpha}$ is chosen so that
$\Pi_{\bfB}(\calC_{1-\alpha}\mid \bfA)=1-\alpha$. Using MCMC draws, we estimate it by the empirical
quantile
\[
r_{1-\alpha} := \mathrm{Quantile}_{1-\alpha}\big(d(\BhatF,\bfB^{(1)}),\ldots,d(\BhatF,\bfB^{(M)})\big).
\]

For local approximations, we work on the smooth rank-$r$ stratum $\calMnro$. Let $\hat{\bfY}$ be
the factor representative so that
$\BhatF=\hat{\bfY}\hat{\bfY}^\top$. For each draw $\bfY^{(m)}$, define the tangent residual
\[
\bfxi^{(m)} := \Log_{\hat{\bfY}}(\bfY^{(m)})\in\calH_{\hat{\bfY}}.
\]
On $\calMnro$, the log lift is compatible with the quotient distance:
$\|\bfxi^{(m)}\|_F=d(\BhatF,\bfB^{(m)})$. Thus the residuals
$\{\bfxi^{(m)}\}$ provide a symmetry-respecting local coordinate system for posterior variation.

We define the empirical tangent covariance operator by
\begin{equation}\label{eq:tangent-cov}
\widehat{\bfSigma}
:=
\frac{1}{M-1}\sum_{m=1}^M \bfxi^{(m)}\otimes \bfxi^{(m)}.
\end{equation}
Its effective dimension is that of the rank-$r$ stratum,
$\dim(\calMnro)=nr-\frac{r(r+1)}{2}$. Diagonalizing $\widehat{\bfSigma}$ yields dominant
directions of local posterior variation, which can be mapped back to the manifold, producing principal geodesic-type visualizations
\citep{fletcher_2004_PrincipalGeodesicAnalysis}.

\subsection{Pairwise distance inference}\label{subsec:dist-infer}

Pairwise distances are the basic inferential targets in LSMs because they determine edge
probabilities in \eqref{eq:lsm}. For any $\bfB\in\calMnr$, define
\[
D_{ij}^2(\bfB) := \bfB_{ii}+\bfB_{jj}-2\bfB_{ij},\qquad D_{ij}(\bfB):=\sqrt{D_{ij}^2(\bfB)}.
\]
These quantities are intrinsic because they depend only on $\bfB$.

The most direct approach is Monte Carlo inference. For each pair $(i,j)$, compute
$D_{ij}^{(m)}:=D_{ij}(\bfB^{(m)})$ and summarize the empirical distribution
$\{D_{ij}^{(m)}\}_{m=1}^M$ by posterior means, medians, credible intervals, or tail probabilities.

When the number of dyads is large, it can be useful to approximate posterior variances using local
linearization. Let $\BhatF$ be the sample \Frechet mean and let $E$ denote a tangential symmetric
perturbation at $\BhatF$. The differential of $D_{ij}^2$ at $\BhatF$ is
\[
\mathrm{d}D_{ij}^2[E]=E_{ii}+E_{jj}-2E_{ij}.
\]
If $D_{ij}(\BhatF)>0$, the delta method gives
\[
\mathrm{d}D_{ij}[E]=\frac{E_{ii}+E_{jj}-2E_{ij}}{2D_{ij}(\BhatF)},
\]
and hence
\[
\mathrm{Var}\{D_{ij}(\bfB)\}\approx \nabla D_{ij}(\BhatF)^\top \widehat{\bfSigma}\,\nabla D_{ij}(\BhatF),
\]
where $\widehat{\bfSigma}$ is the tangent covariance operator \eqref{eq:tangent-cov}. If
$D_{ij}(\BhatF)=0$, it is more stable to apply the delta method to $D_{ij}^2$, which is
differentiable everywhere.

\subsection{Inference for edge probabilities and link-scale effects}\label{subsec:edgeprob}

Distances enter the likelihood through the link-scale term
$\alpha-\|\bfx_i-\bfx_j\|=\alpha-D_{ij}(\bfB)$. Intrinsic posterior inference for edge probabilities
therefore propagates the posterior of $(\alpha,\bfB)$ through
\[
p_{ij}(\alpha,\bfB):=g\!\left(\alpha-D_{ij}(\bfB)\right).
\]
With posterior draws $\{(\alpha^{(m)},\bfB^{(m)})\}_{m=1}^M$, we compute
\[
p_{ij}^{(m)} := g\!\left(\alpha^{(m)}-D_{ij}(\bfB^{(m)})\right).
\]
Credible intervals for $p_{ij}$, as well as summaries for the link-scale effect
$\alpha-D_{ij}$, then follow directly from the empirical distribution.

The same quantities support posterior predictive checks. Given $p_{ij}^{(m)}$, we may simulate
replicate networks $\bfA^{\mathrm{rep},(m)}$ by drawing edges independently with probabilities
$\{p_{ij}^{(m)}\}$ and compare relevant network summaries between $\bfA$ and
$\{\bfA^{\mathrm{rep},(m)}\}$. This preserves identifiability because the predictive mechanism
depends only on $(\alpha,\bfB)$.

\subsection{Node-level uncertainty summaries}\label{subsec:node-uncert}

Beyond pairwise inference, it is often useful to summarize uncertainty at the node level. Because
latent coordinates are not identifiable, such summaries should depend only on invariant quantities.
We therefore aggregate posterior variability in distances involving each node. For node $i$, define
\begin{equation}\label{eq:Ui}
U_i := \frac{1}{n-1}\sum_{j\neq i}\mathrm{Var}\{D_{ij}(\bfB)\}.
\end{equation}
Here $\mathrm{Var}\{D_{ij}(\bfB)\}$ may be estimated either by the empirical variance of
$\{D_{ij}^{(m)}\}_{m=1}^M$ or by the local approximation based on $\widehat{\bfSigma}$. The index
$U_i$ measures posterior uncertainty in node $i$'s relational position and can highlight regions of
the network where latent structure is weakly identified.

\subsection{Visualization after inference}\label{subsec:viz}

Visualization is often central to the interpretation of LSMs, but it should be separated from
inference because coordinate orientation is not identifiable. Once intrinsic summaries such as
$\BhatF$, credible sets, and $\{U_i\}$ have been computed, one may choose any convenient
representative embedding for display.

To visualize the posterior mean structure, compute an eigendecomposition
$\BhatF=\bfU\bfLambda\bfU^\top$ and set
$
\hat{\bfX}:=\bfU_{[:,1:r]}\bfLambda_{1:r}^{1/2},
$
so that $\hat{\bfX}\hat{\bfX}^\top=\BhatF$ up to numerical error. Any rotated embedding
$\hat{\bfX}\bfR$ for $\bfR\in O(r)$ yields the same distances and is therefore equally valid for
visualization; the choice of $\bfR$ affects appearance only.

To depict posterior variability qualitatively, one may align each factor $\bfY^{(m)}$ to the mean
factor $\hat{\bfY}$ via Procrustes and plot the aligned coordinates
$\bfY^{(m)}\bfR_m^\star$. This is a display convention built on an already defined intrinsic mean
and does not alter the posterior summaries themselves.

Node-level uncertainty indices can be incorporated directly into visualizations, for example by
scaling point size or transparency according to $\sqrt{U_i}$. Such displays convey uncertainty in
relational position without relying on coordinate-wise credible regions, which are not meaningful
under rotational non-identifiability.

\subsection{Computational considerations}\label{subsec:comp}

The computational cost of the proposed summaries is driven mainly by repeated low-rank operations on
the factors $\bfY^{(m)}$. A single evaluation of $d(\bfB_1,\bfB_2)$ requires forming
$\bfY_2^\top\bfY_1$ and computing an SVD of an $r\times r$ matrix, with costs $O(nr^2)$ and
$O(r^3)$, respectively. Since $r$ is typically small, the dominant cost is $O(nr^2)$.

A full iteration costs $O(Mnr^2+Mr^3)$, plus the $O(r^3)$ cost of
solving the Lyapunov equation \eqref{eq:sylvester}. Memory usage is dominated by storing
$\{\bfY^{(m)}\}$ rather than $\{\bfB^{(m)}\}$, requiring $O(Mnr)$ rather than $O(Mn^2)$ memory.

In practice, centering should be enforced on factors by applying $\bfH$ to mitigate numerical
drift. When Gram matrices are formed explicitly, symmetry can be enforced by replacing $\bfB$ with
$(\bfB+\bfB^\top)/2$. If posterior draws approach the rank-deficient boundary, one may regularize
computations by truncating very small eigenvalues or restricting local linearization to draws that
remain in the rank-$r$ stratum.

\section{Theory}\label{sec:theory}

We now establish basic properties of the quotient-based posterior summaries introduced above. Let
$\Pi(\cdot\mid \bfA)$ denote the posterior distribution on the latent configuration $\bfX$, and let
$\Pi_{\bfB}(\cdot\mid \bfA)$ be its pushforward through the centered Gram map $\Phi$, that is, the
induced posterior distribution of $\bfB=\Phi(\bfX)$ supported on $\calMnr$. We write
$(\calMnr,d)$ for the metric space defined by \eqref{eq:state-space} and
\eqref{eq:quotient-distance}. To distinguish population-level objects from their empirical
approximations based on MCMC output, we use $\BF$ for a population \Frechet mean of the induced
posterior and reserve $\BhatF$ for the sample \Frechet mean from
Section~\ref{sec:summaries}.

\subsection{Canonicality}\label{subsec:canon}

We call a posterior summary canonical if it is a measurable functional of
$\Pi_{\bfB}(\cdot\mid \bfA)$. Since the map $\Phi$ removes the rigid-motion
non-identifiabilities, any measurable functional of the induced posterior on $\calMnr$ is
automatically invariant to those symmetries.

\begin{theorem}[Canonicality]\label{thm:canonical}
Any measurable functional of $\Pi_{\bfB}(\cdot\mid \bfA)$ is invariant under rigid motions of the
latent coordinates and hence is canonical. In particular, the population \Frechet mean
\[
\BF\in\argmin_{\bfB\in\calMnr}\int d^2(\bfB,\bfB')\,\Pi_{\bfB}(d\bfB'\mid\bfA),
\]
the corresponding \Frechet variation, intrinsic credible balls, distance-based summaries,
probability-based summaries, and node-level uncertainty indices are canonical.
\end{theorem}


The empirical analogues obtained from MCMC output are canonical with respect to the corresponding
empirical induced posterior. In many Bayesian LSMs, translation is not a posterior symmetry because
the prior anchors the configuration near the origin, whereas orthogonal invariance typically
persists; it is this remaining nonidentifiability that the quotient representation resolves.

\subsection{Existence of the posterior \Frechet mean}\label{subsec:exist}

Since $\calMnr$ allows rank deficiency, it is closed. This avoids the boundary pathology that would
arise if one minimized only over the rank-exact stratum $\calMnro$. The next result shows that the
induced posterior admits a \Frechet mean under a finite second-moment condition.

\begin{theorem}[Existence]\label{thm:existence}
Assume that $\Pi_{\bfB}(\cdot\mid \bfA)$ has finite second moment with respect to $d$, that is,
\[
\int d^2(\bfB,\bfB_0)\,\Pi_{\bfB}(d\bfB\mid \bfA)<\infty
\qquad\text{for some }\bfB_0\in\calMnr.
\]
Then there exists at least one \Frechet mean
\[
\BF\in\argmin_{\bfB\in\calMnr}
\int d^2(\bfB,\bfB')\,\Pi_{\bfB}(d\bfB'\mid \bfA).
\]
\end{theorem}

Theorem~\ref{thm:existence} guarantees existence on the closed space $\calMnr$, but not necessarily
rank exactly $r$. If posterior mass accumulates near the rank-deficient boundary, a \Frechet mean
may also lie there. A simple sufficient condition for $\BF\in\calMnro$ is that the posterior be
supported away from the boundary, e.g., if $\lambda_r(\bfB)\ge \epsilon>0$ almost surely for
some $\epsilon>0$.

\subsection{Stability under perturbations of the posterior}\label{subsec:stability}

A coherent posterior summary should vary continuously under small perturbations of the posterior law.
This matters conceptually, and also practically because exact posteriors are typically approximated
by finite MCMC output.

Let $\calP_2(\calMnr)$ denote the set of probability measures on $\calMnr$ with finite second
moment under $d$. For $P\in\calP_2(\calMnr)$, define
\[
F_P(\bfB):=\int d^2(\bfB,\bfB')\,P(d\bfB'),\qquad \bfB\in\calMnr,
\]
and let $\mu_P$ denote a \Frechet mean of $P$. Suppose $\mu_P$ is unique and $F_P$ satisfies the
local quadratic growth condition
\begin{equation}\label{eq:qgrowth}
F_P(\bfB)\ge F_P(\mu_P)+\frac{\lambda}{2}\,d^2(\bfB,\mu_P)
\qquad \text{for all } \bfB\in\calMnr
\end{equation}
for some $\lambda>0$.

\begin{theorem}[Stability under Wasserstein perturbations]\label{thm:stability}
Let $P,Q\in\calP_2(\calMnr)$. Assume that $\mu_P$ is unique and that the quadratic growth condition
\eqref{eq:qgrowth} holds with constant $\lambda>0$. Let $\mu_Q$ be any \Frechet mean of $Q$, and let
\[
w:=W_2(P,Q)
\]
denote the $2$-Wasserstein distance on $\calP_2(\calMnr)$ induced by $d$. Then
\[
d(\mu_P,\mu_Q)
\le
\frac{4\sqrt{2}}{\lambda}\,w
+
2^{5/4}\sqrt{\frac{w}{\lambda}}\,
\Big(\sqrt{F_P(\mu_P)}+\sqrt{F_Q(\mu_Q)}\Big)^{1/2}.
\]
In particular, $d(\mu_P,\mu_Q)\to 0$ as $W_2(P,Q)\to 0$.
\end{theorem}

Thus the posterior mean depends continuously on the posterior law under a quantitative regularity
condition. In particular, if $P$ is the exact induced posterior and $Q$ is an empirical or
approximate version, closeness in Wasserstein distance implies closeness of the corresponding
\Frechet means. This underlies the stability of the sample-based procedures in
Section~\ref{sec:summaries}.

\subsection{Consistency of distance-based inference}\label{subsec:consistency}

We now consider asymptotic behavior as the amount of information in the observed network increases.
Let $\bfX^\star$ denote a target latent configuration and let
$
\bfB^\star := \Phi(\bfX^\star)
$
be the corresponding identifiable latent structure. Suppressing sample-size indexing, assume that
the induced posterior concentrates at $\bfB^\star$ in mean-square quotient distance:
\begin{equation}\label{eq:posterior-conc}
\int d^2(\bfB,\bfB^\star)\,\Pi_{\bfB}(d\bfB\mid \bfA)\xrightarrow{P}0.
\end{equation}
This condition is stronger than weak posterior concentration and is exactly what is needed to
control posterior means and variances of intrinsic functionals.

Under \eqref{eq:posterior-conc}, any population posterior \Frechet mean concentrates at the true
identifiable structure, and hence so do continuous distance-based summaries. The next result is
stated for a population posterior \Frechet mean $\BF$ induced by $\Pi_{\bfB}(\cdot\mid\bfA)$. A
corresponding statement for the empirical summary $\BhatF$ requires an additional approximation step
controlling the MCMC error, obtainable by combining posterior consistency with
Theorem~\ref{thm:stability}.

\begin{theorem}[Consistency of pairwise distances]\label{thm:dist-cons}
Assume \eqref{eq:posterior-conc}. Then for any fixed pair $(i,j)$,
\[
D_{ij}(\BF) \xrightarrow{P} D_{ij}(\bfB^\star).
\]
\end{theorem}

Hence, under posterior mean-square concentration on the quotient space, any population posterior
\Frechet mean recovers the true pairwise geometry asymptotically. This is the relevant notion of
consistency for LSMs as latent coordinates are defined only up to rigid motions.

\begin{corollary}[Consistency of edge probabilities]\label{cor:prob-cons}
Assume \eqref{eq:posterior-conc}. Suppose also that $\bar{\alpha}$ is a consistent point summary of
the marginal posterior of $\alpha$, so that $\bar{\alpha}\xrightarrow{P}\alpha^\star$, and that the
link function $g$ is continuous. Then for each fixed pair $(i,j)$,
\[
g\!\left(\bar{\alpha}-D_{ij}(\BF)\right)
\xrightarrow{P}
g\!\left(\alpha^\star-D_{ij}(\bfB^\star)\right).
\]
\end{corollary}

\subsection{Asymptotic behavior of node-level uncertainty}\label{subsec:node-asymp}

The node-level uncertainty indices quantify posterior uncertainty about each node's relational
position. Under posterior concentration, such uncertainty should vanish.

\begin{theorem}[Vanishing node-level uncertainty]\label{thm:node-vanish}
Assume \eqref{eq:posterior-conc}. Then for each fixed node $i$,
\[
U_i\xrightarrow{P}0.
\]
\end{theorem}

\section{Experiments}\label{sec:experiments}

In all experiments, posterior draws were generated with the \texttt{latentnet} package
\citep{krivitsky_2008_FittingPositionLatent} in \textsf{R}
\citep{rcoreteam_2025_LanguageEnvironmentStatistical}, and the proposed quotient-based summaries were applied as post-processing. The aim is to assess the inferential value of these summaries rather than to propose a new fitting algorithm.

\subsection{Simulation study}\label{subsec:sim}

We considered a two-dimensional LSM with $n=120$ nodes partitioned into three latent groups,
$c_i\in\{L,B,R\}$, corresponding to a left core, a bridge group, and a right core, with sizes
$n_L=48$, $n_B=24$, and $n_R=48$. Conditional on group membership, latent positions were generated as
\[
\bfx_i^\star \mid (c_i=g)\sim \mathcal N(\mu_g,\sigma_g^2\bfI_2),
\qquad g\in\{L,B,R\}.
\]
Let $\bfX^\star=(\bfx_1^\star,\ldots,\bfx_n^\star)^\top\in\bbR^{n\times 2} $
denote the resulting latent configuration, and let $\bfB^\star=\Phi(\bfX^\star)=\bfH\bfX^\star\bfX^{\star\top}\bfH$ be the corresponding identifiable target.

We studied two regimes. In the {well-identified} regime, the two core groups were well
separated and the bridge group was concentrated between them with mean $\mu_L=(-1.8,0)$, $\mu_B=(0,0.9)$, $\mu_R=(1.8,0)$ and variance  $\sigma_L=\sigma_R=0.20$ and $\sigma_B=0.25$ In the {weakly identified} regime, the two core groups were closer together and the bridge group
was more diffuse via mean $\mu_L=(-1.25,0)$, $\mu_B=(0,0)$,$\mu_R=(1.25,0)$ and variance $\sigma_L=\sigma_R=0.20$ and $\sigma_B=0.45$. The second regime therefore produces a substantially more difficult posterior summarization problem.

Conditional on $\bfX^\star$, edges were generated independently according to
\[
\bbP(A_{ij}=1\mid \bfX^\star,\alpha^\star)
=
\operatorname{logit}^{-1}\!\left(\alpha^\star-\|\bfx_i^\star-\bfx_j^\star\|\right),
\qquad 1\le i<j\le n.
\]
The intercept $\alpha^\star$ was chosen to yield expected edge density $0.1$. For each regime, we
generated $10$ independent latent templates and $5$ graphs per template, for a total of $50$
networks. Then, each network was fit with the correctly specified two-dimensional LSM. After a burn-in of
$10{,}000$ iterations and thinning by $20$, we retained $M=500$ posterior draws $\{(\bfX^{(m)},\alpha^{(m)})\}_{m=1}^M$
from which we formed
\[
\bfY^{(m)}=\bfH\bfX^{(m)},\qquad
\bfB^{(m)}=\bfY^{(m)}\bfY^{(m)\top}.
\]
We then computed the quotient \Frechet mean $\BhatF$ and two fixed-reference Procrustes means, using
the first posterior draw and the quotient medoid draw as references. To assess reference
sensitivity, we additionally randomized the orientation of each centered draw before forming the
Procrustes summaries; this leaves $\bfB^{(m)}$ unchanged and hence preserves the induced posterior
on the quotient space.

\begin{figure}[t]
\centering
\includegraphics[width=\textwidth]{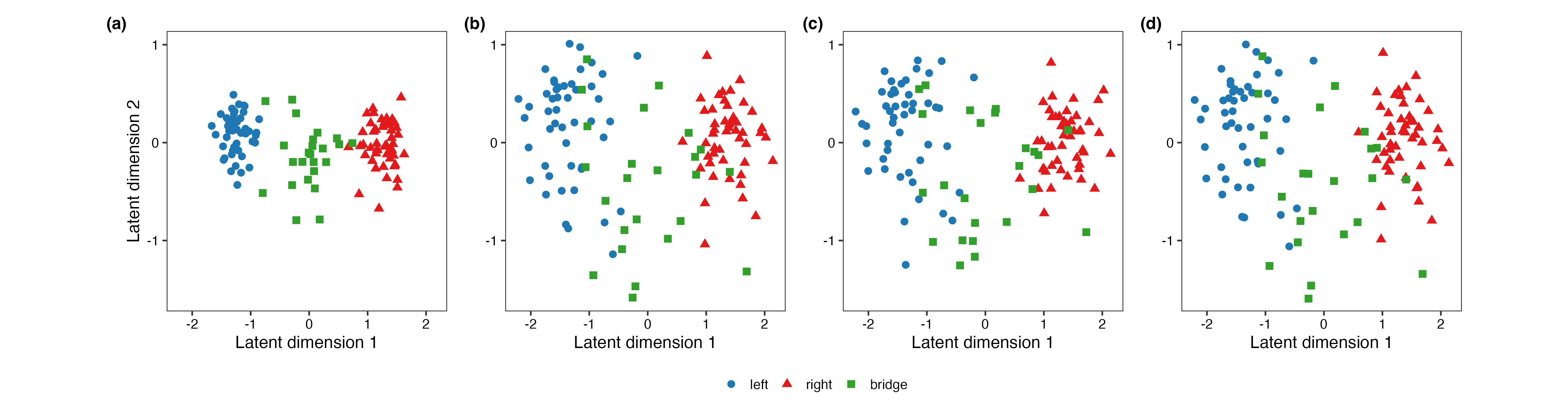}
\caption{Representative dataset from the weakly identified regime. (a) True centered latent
positions. (b) Quotient \Frechet mean. (c) Procrustes mean using the first posterior draw as
reference. (d) Procrustes mean using the quotient medoid draw as reference. Panels (b)--(d) are
aligned to the true centered configuration for display only.}
\label{fig:sim-representative}
\end{figure}

Figure~\ref{fig:sim-representative} shows a representative weak-regime dataset. The quotient
\Frechet mean preserves the three-group geometry, with the bridge nodes occupying an intermediate
and more diffuse region. The Procrustes mean based on the first draw is visibly distorted,
particularly for the left and bridge groups. The medoid-reference summary performs better, but it
remains tied to an external convention.

To quantify dependence on the choice of reference, let $\widehat{\bfB}^{(k)}_{\mathrm{Proc}}$ be the
Gram matrix induced by the fixed-reference Procrustes mean under reference $k$, and define
\[
S_{\mathrm{ref}}
=
\frac{2}{K(K-1)}
\sum_{1\le k<\ell\le K}
\left\|
\widehat{\bfB}^{(k)}_{\mathrm{Proc}}-
\widehat{\bfB}^{(\ell)}_{\mathrm{Proc}}
\right\|_F,
\qquad K=10.
\]

\begin{figure}[t]
\centering
\includegraphics[width=\textwidth]{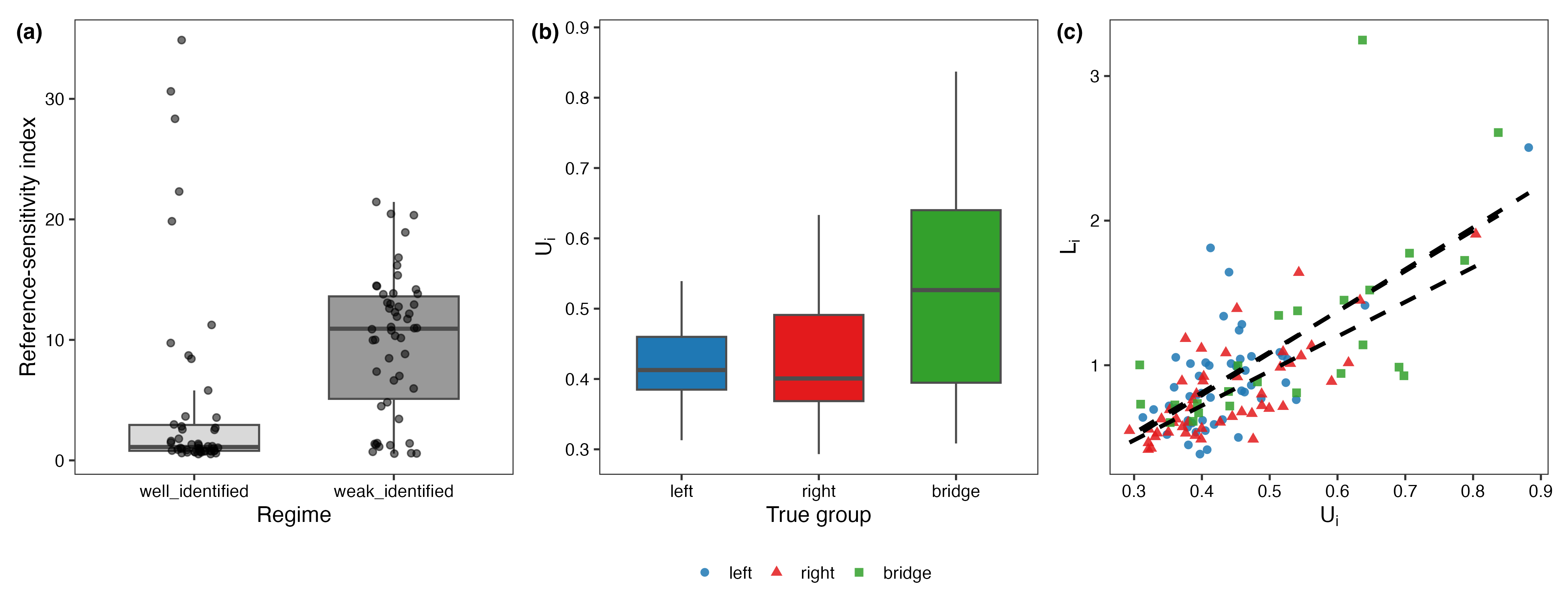}
\caption{Simulation summaries. (a) Reference-sensitivity index $S_{\mathrm{ref}}$ by regime.
(b) Node-level uncertainty index $U_i$ by true group for the representative weak-regime dataset.
(c) Relationship between $U_i$ and the invariant node-wise loss $L_i$.}
\label{fig:sim-uncertainty}
\end{figure}

Figure~\ref{fig:sim-uncertainty}(a) shows that $S_{\mathrm{ref}}$ is small in the well-identified
regime but much larger in the weakly identified regime, confirming that alignment-based summaries
become increasingly reference-sensitive as posterior concentration weakens.

We also examined the node-level uncertainty index
\[
U_i=\frac{1}{n-1}\sum_{j\neq i}\mathrm{Var}\{D_{ij}(\bfB)\},
\]
together with the invariant node-wise loss
\[
L_i
=
\frac{1}{n-1}\sum_{j\neq i}
\left\{\bar D_{ij}-D_{ij}(\bfB^\star)\right\}^2,
\qquad
\bar D_{ij}=\frac{1}{M}\sum_{m=1}^M D_{ij}(\bfB^{(m)}).
\]

Figure~\ref{fig:sim-uncertainty}(b) shows that bridge nodes have the largest uncertainty, as
expected from the data-generating design, and panel~(c) shows a clear positive association between
$U_i$ and $L_i$. Additional quantitative diagnostics are reported in the Supplementary Material.

Overall, the simulation shows that quotient-based summaries remain well defined when
reference-based Procrustes summaries become unstable, and that the proposed uncertainty index
aligns with actual estimation error.

\subsection{Florentine marriage network}\label{subsec:florentine}

We next analyzed the Florentine marriage network, a standard benchmark in latent space modeling
\citep{hoff_2002_LatentSpaceApproaches}. We fit a two-dimensional LSM  to the $16\times 16$
undirected marriage adjacency matrix. To remain close to the setup in
\citet{hoff_2002_LatentSpaceApproaches}, we used a Gaussian working prior on the intercept with mean
$2$ and variance $4$, together with a diffuse Gaussian prior on the latent coordinates centered at
the origin. The Florentine business network was reserved for interpretation and was not used in
fitting.

\begin{figure}[ht]
\centering
\includegraphics[width=.8\textwidth]{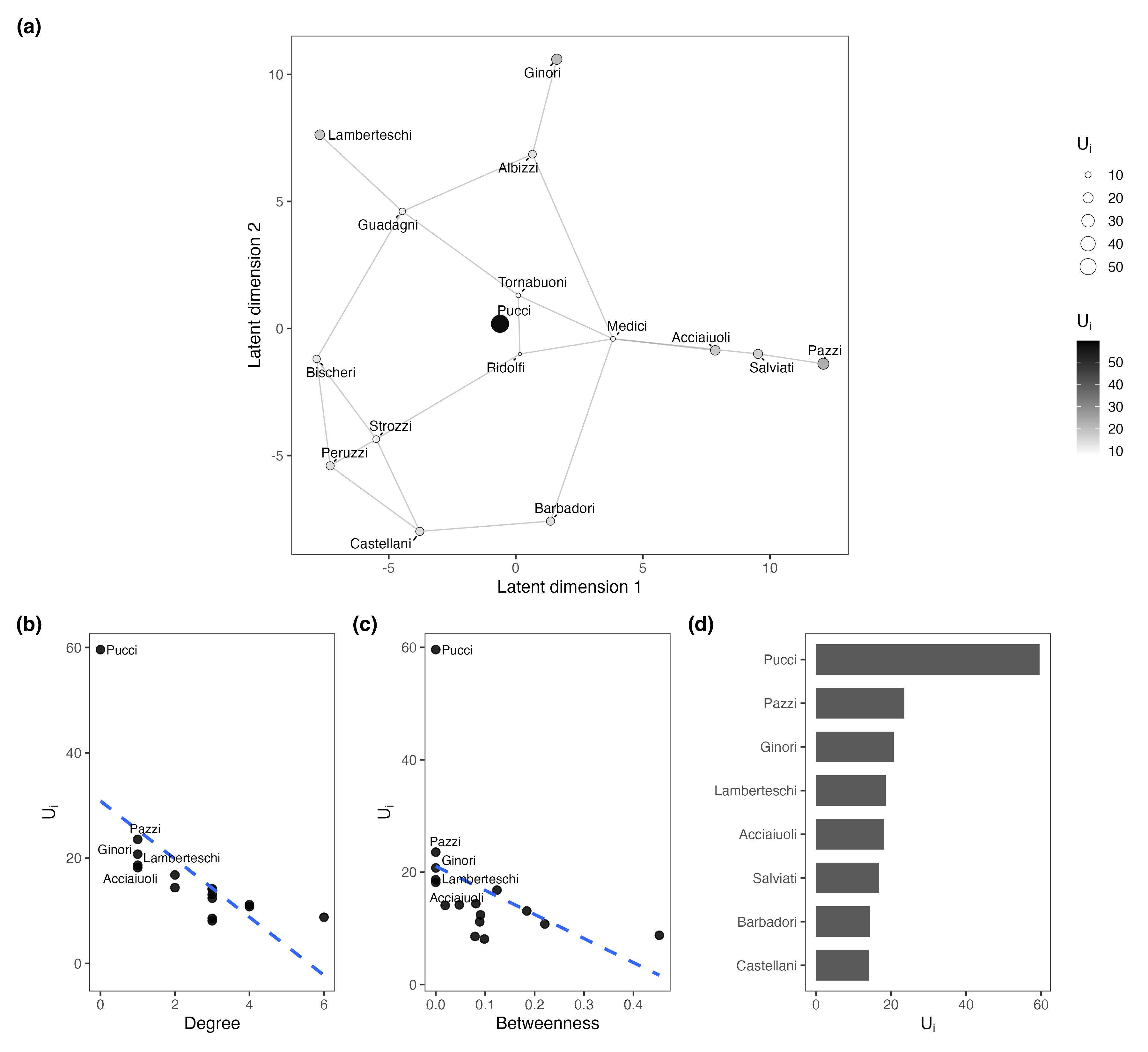}
\caption{Posterior summaries for the Florentine marriage network. (a) Quotient \Frechet mean
embedding with observed marriage edges overlaid; point size and shade are proportional to
$U_i$. (b) Node-level uncertainty $U_i$ versus degree. (c) Node-level uncertainty $U_i$ versus
betweenness. (d) Families with the largest posterior uncertainty.}
\label{fig:florentine-main}
\end{figure}

Figure~\ref{fig:florentine-main} summarizes the posterior at the node level. Panel~(a) shows the
quotient \Frechet mean embedding with observed marriages overlaid and node size proportional to the
uncertainty index $U_i$. The Medici family occupies a central position, but the uncertainty display
reveals information not visible from the mean embedding alone. In particular, Pucci has by far the
largest uncertainty despite lying near the center of the posterior mean configuration. Families such
as Pazzi, Ginori, Lamberteschi, and Acciaiuoli also exhibit elevated uncertainty. Panels~(b)--(d)
show that $U_i$ tends to decrease with degree and betweenness, although these associations are far
from deterministic. Thus $U_i$ is not merely a restatement of standard centrality measures; it
captures posterior ambiguity in relational position.

\begin{figure}[ht]
\centering
\includegraphics[width=\textwidth]{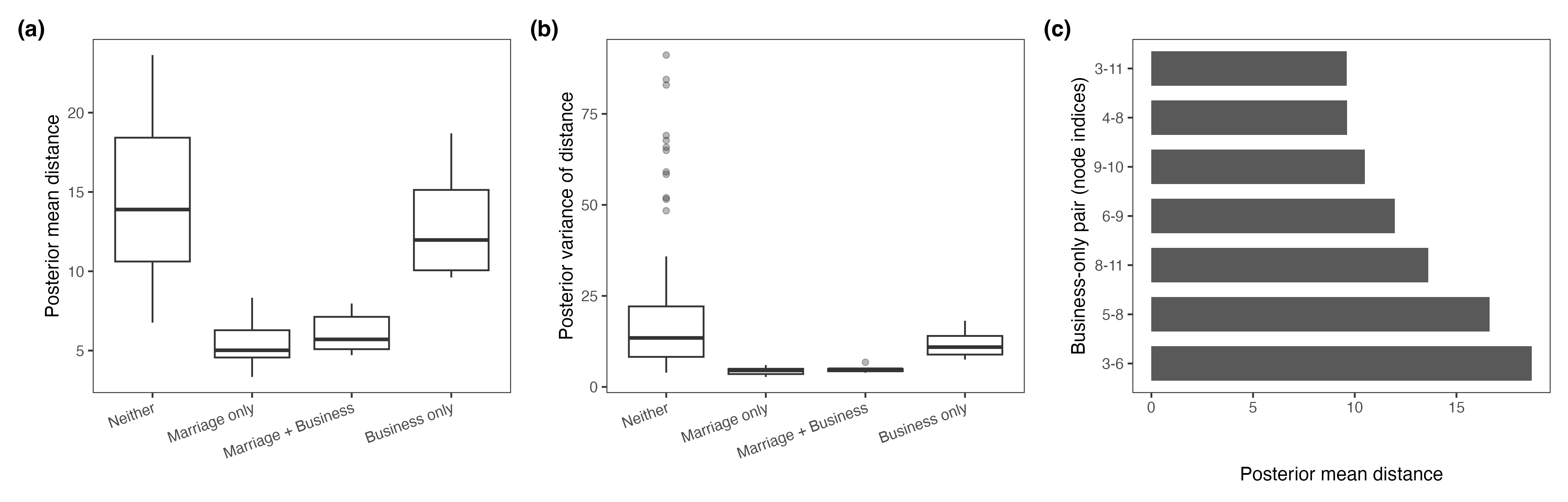}
\caption{Dyad-level posterior summaries for the Florentine marriage network. (a) Posterior mean
distance by dyad type. (b) Posterior variance of distance by dyad type. (c) Business-only dyads
with the smallest posterior mean distances. The business network was not used in fitting and is used
here only for interpretation.}
\label{fig:florentine-dyad}
\end{figure}

Figure~\ref{fig:florentine-dyad} examines dyad-level summaries using the business network as an
external benchmark. Marriage-linked dyads have the smallest posterior mean distances and the
smallest posterior distance variances. Business-only dyads occupy an intermediate range, lying
closer than dyads with neither tie but farther than marriage-linked pairs. Thus the latent geometry
learned from the marriage network also reflects economic structure not used in fitting. The
business-only dyads with the smallest posterior mean distances are natural candidates for latent
relationships supported by the fitted model even though they do not appear as marriage edges.

Overall, the Florentine example shows that a single embedding captures only posterior central
tendency, whereas quotient-based summaries distinguish stable from ambiguous family positions and
reveal latent affinities not apparent from aligned point estimates alone.

\subsection{Coauthorship network}\label{subsec:coauthorship}

The final example considers a coauthorship network of 4{,}383 statisticians
\citep{ji_2022_CocitationCoauthorshipNetworks}. Following the preprocessing in that work, an
undirected edge is placed between two authors if they have coauthored at least three papers. We
focus on three first-layer biostatistics communities, labeled \emph{Biostatistics (Europe)},
\emph{Biostatistics (UNC)}, and \emph{Biostatistics (Michigan)}, with sizes 202, 673, and 264,
respectively, for a total of 1{,}139 authors. We then restrict attention to the induced subgraph on
these authors, retain its largest connected component of size 980, and for computational
feasibility fit the model on a connected 400-node subnetwork. Preprocessing details are reported in
the Supplementary Material.

\begin{figure}[ht]
\centering
\includegraphics[width=.9\textwidth]{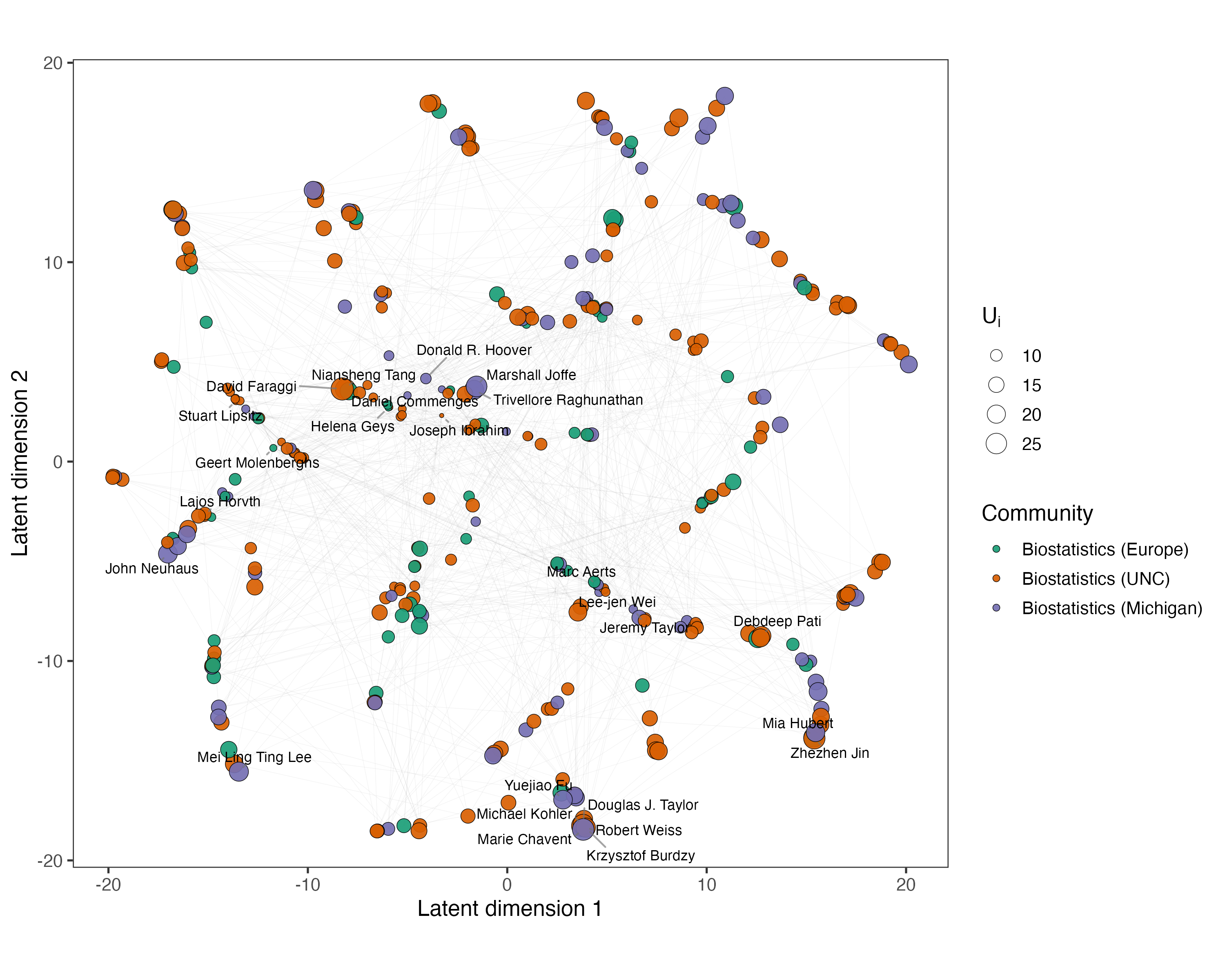}
\caption{Quotient \Frechet mean embedding for the coauthorship network, with observed coauthorship
edges overlaid. Colors indicate the three biostatistics communities, and node size is proportional
to the node-level uncertainty index $U_i$.}
\label{fig:coauthorship-embed}
\end{figure}

Figure~\ref{fig:coauthorship-embed} shows the quotient \Frechet mean embedding with observed
coauthorship edges overlaid. The three communities are not sharply separated in the posterior mean
geometry. Instead, authors from the three groups are substantially intermixed, indicating that the
dominant latent variation in the fitted network is not explained by the community labels alone. The
uncertainty display adds important information: authors such as David Faraggi, Michael Kohler,
Marie Chavent, Zhezhen Jin, and Marshall Joffe have relatively large $U_i$, but they are not
confined to a single part of the embedding. Rather, they occupy regions where multiple latent
placements are compatible with the observed network.

\begin{figure}[ht]
\centering
\includegraphics[width=\textwidth]{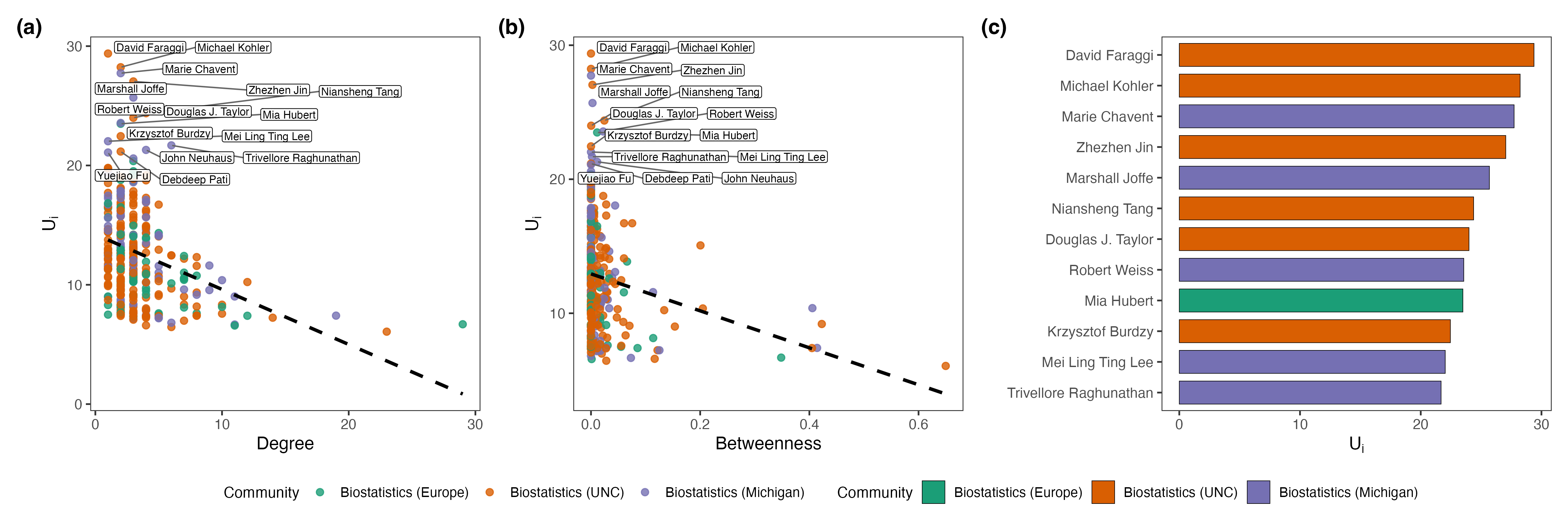}
\caption{Node-level posterior uncertainty diagnostics for the coauthorship network.
(a) $U_i$ versus degree.
(b) $U_i$ versus betweenness.
(c) Authors with the largest posterior uncertainty.}
\label{fig:coauthorship-diagnostics}
\end{figure}

Figure~\ref{fig:coauthorship-diagnostics} relates $U_i$ to standard graph summaries. Uncertainty
shows an overall negative association with degree and betweenness, but these relationships are again
far from deterministic. Authors with similar graph summaries can have markedly different posterior
uncertainty, so $U_i$ captures posterior ambiguity in relational position rather than merely
restating graph centrality. The most uncertain authors are spread across all three communities,
although the UNC group is somewhat more prominent among the most uncertain nodes in this fitted
subnetwork.

\begin{figure}[ht]
\centering
\includegraphics[width=\textwidth]{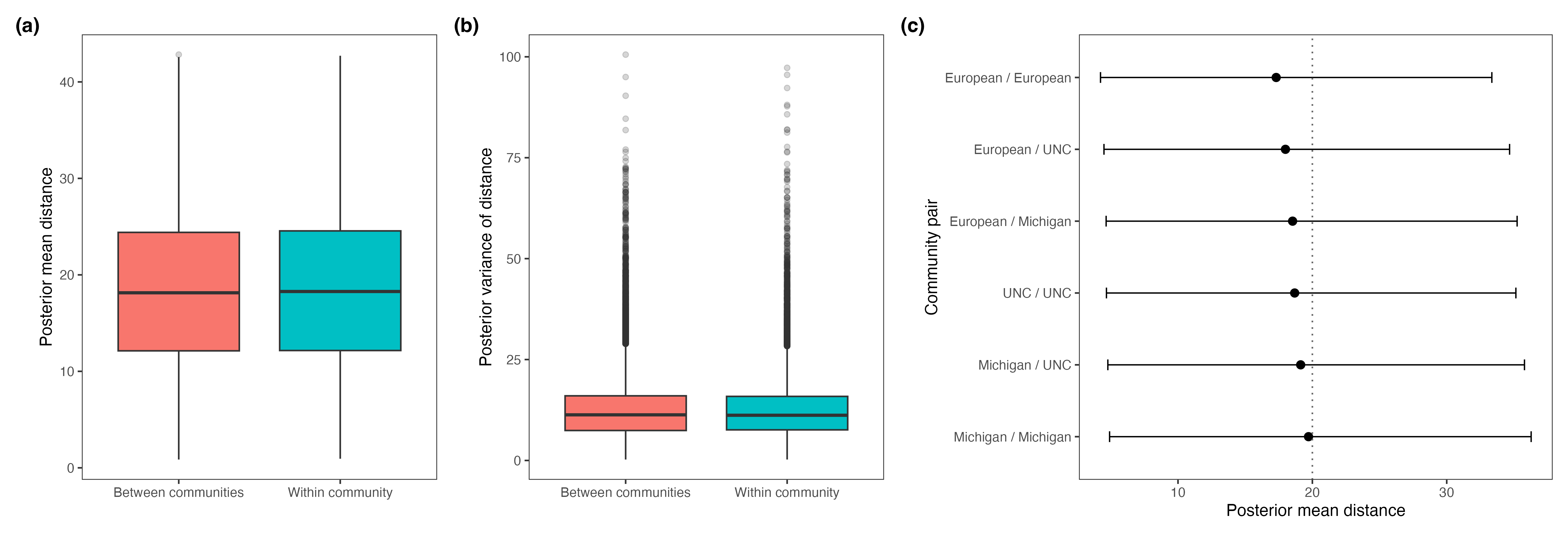}
\caption{Dyad-level posterior summaries for the coauthorship network.
(a) Posterior mean distances for within-community and between-community dyads.
(b) Posterior variances of distances for within-community and between-community dyads.
(c) Posterior mean distances with uncertainty intervals for each unordered community pair.}
\label{fig:coauthorship-dyads}
\end{figure}

Dyad-level summaries are presented in Figure~\ref{fig:coauthorship-dyads}. Within-community and
between-community dyads have very similar posterior mean distances and posterior distance variances.
Panel~(c) refines this by comparing all community pairs and shows substantial overlap of the
corresponding uncertainty intervals. Taken together, these displays indicate that the three selected
communities are porous rather than sharply separated collaboration groups.

This example complements the Florentine analysis by showing that quotient-based uncertainty remains
interpretable in a substantially larger network. Here the main message is not that the selected
communities form clean latent clusters, but that their collaboration patterns generate overlap and
boundary ambiguity, which become visible through node-level and dyad-level posterior uncertainty.

\section{Conclusion}\label{sec:conclusion}

This paper argues that posterior analysis for LSMs should be carried out on
identifiable latent structure rather than on arbitrary coordinate representatives. The centered Gram
map provides a natural quotient representation for this purpose: it removes rigid-motion
nonidentifiability while preserving exactly the pairwise geometry that drives the likelihood. This
viewpoint leads to canonical posterior summaries that depend only on the induced posterior
distribution on the quotient space, not on an external alignment convention.

Within this framework, we introduced an intrinsic posterior mean through a \Frechet criterion,
together with global dispersion, intrinsic credible regions, dyad-level summaries for distances and
edge probabilities, and node-level uncertainty indices. These quantities are computable directly
from MCMC output using only small orthogonal Procrustes problems and low-rank matrix operations. We
also established basic theoretical guarantees, including canonicality, existence of \Frechet means
on a closed state space, stability under perturbations of the posterior law, and consistency of
distance-based inference under posterior concentration.

The empirical results reinforce the main methodological message. In the simulation study,
alignment-based summaries were broadly similar to quotient-based summaries when the posterior was
well concentrated, but became increasingly reference-dependent as identification weakened. The
Florentine marriage network showed that quotient-based uncertainty can reveal ambiguous family
positions and latent affinities not evident from a single embedding alone. The coauthorship example
demonstrated that the same framework remains informative in a larger contemporary network, where
node-level and dyad-level uncertainty helped clarify porous community boundaries and boundary-spanning
collaboration patterns.

Several extensions merit further study. One direction is to adapt the framework to directed,
weighted, bipartite, or dynamic LSMs. Another is to develop analogous quotient-based
posterior summaries for non-Euclidean latent spaces such as hyperbolic or spherical models. On the
computational side, large-scale approximations and streaming implementations would be valuable.
Because the present framework operates as a post-processing layer on posterior draws, improving its
scalability would broaden the practical reach of coherent uncertainty quantification in latent
network models.

The main practical conclusion is simple: alignment remains useful for visualization, but it should
follow inference rather than define it. By separating inferential content from arbitrary
orientation, the quotient perspective yields posterior summaries that are both interpretable and
faithful to the symmetry structure of the model.





\bibliographystyle{dcu}
\bibliography{references}

\newpage
\section*{Supplementary Materials}\label{sec:supp}

\subsection*{Proofs}
\subsubsection*{Proof of Lemma 1}
If $\tilde \bfX=\tilde \bfW \bfR$ with $\bfR\in O(r)$, then
\[
\tilde \bfX\tilde \bfX^\top=\tilde \bfW \bfR\bfR^\top\tilde \bfW^\top=\tilde \bfW\tilde \bfW^\top.
\]
Conversely, if $\tilde \bfX\tilde \bfX^\top=\tilde \bfW\tilde \bfW^\top$ and both have full column
rank $r$, then there exists an invertible $\bfQ$ with $\tilde \bfX=\tilde \bfW \bfQ$. Plugging into
the Gram equality gives $\bfQ\bfQ^\top=\bfI_r$, hence $\bfQ\in O(r)$.

For the uncentered equivalence, note that $\bfH\bfone_n=\bfzero$, so for any $\bft\in\bbR^r$,
\[
\bfH\big(\bfW\bfR+\bfone_n \bft^\top\big)=\bfH \bfW \bfR.
\]
Thus $\bfX=\bfW\bfR+\bfone_n \bft^\top$ implies $\Phi(\bfX)=\Phi(\bfW)$. Conversely, if
$\Phi(\bfX)=\Phi(\bfW)$ then $\bfH \bfX=\bfH \bfW \bfR$ for some $\bfR\in O(r)$ by the first part,
and therefore $\bfX-\bfW\bfR$ must be a pure translation, i.e.,
$\bfX=\bfW\bfR+\bfone_n \bft^\top$ for some $\bft\in\bbR^r$.
\qed

\subsubsection*{Proof of Theorem 2}
Let $g=(\bfR,\bft)\in O(r)\times\bbR^r$ act on latent configurations by
\[
g\cdot \bfX := \bfX\bfR+\bfone_n\bft^\top.
\]
By construction of the centered Gram map,
\[
\Phi(g\cdot \bfX)
=
\bfH(\bfX\bfR+\bfone_n\bft^\top)(\bfX\bfR+\bfone_n\bft^\top)^\top\bfH.
\]
Since $\bfH\bfone_n=\bfzero$, all terms involving $\bft$ vanish after centering, and we obtain
\[
\Phi(g\cdot \bfX)
=
\bfH\bfX\bfR\bfR^\top\bfX^\top\bfH
=
\bfH\bfX\bfX^\top\bfH
=
\Phi(\bfX).
\]
Therefore, the map $\Phi$ is invariant under rigid motions. It follows that the induced posterior
law of $\bfB=\Phi(\bfX)$ is unchanged if one replaces $\bfX$ by $g\cdot\bfX$. Hence any measurable
functional of $\Pi_{\bfB}(\cdot\mid \bfA)$ is invariant under rigid motions of the latent
coordinates.

The summaries listed in the statement are all measurable functionals of $\Pi_{\bfB}(\cdot\mid \bfA)$
by their definitions. Therefore they are canonical.
\qed

\subsubsection*{Proof  of Theorem 3}
Let
\[
F(\bfB):=\int d^2(\bfB,\bfB')\,\Pi_{\bfB}(d\bfB'\mid \bfA),\qquad \bfB\in\calMnr,
\]
denote the posterior \Frechet functional. By assumption, $F(\bfB_0)<\infty$ for some
$\bfB_0\in\calMnr$. Let $\{\bfB_n\}_{n\ge 1}\subset\calMnr$ be a minimizing sequence, so that
\[
F(\bfB_n)\downarrow \inf_{\bfB\in\calMnr} F(\bfB).
\]
Since $F(\bfB_n)\le F(\bfB_0)+1$ for all sufficiently large $n$, it suffices to work with such $n$.

We first show that $\{\bfB_n\}$ is bounded in the metric $d$. For any fixed $\bfB'\in\calMnr$, the
triangle inequality gives
\[
d(\bfB_n,\bfB_0)\le d(\bfB_n,\bfB')+d(\bfB',\bfB_0).
\]
Integrating with respect to $\Pi_{\bfB}(d\bfB'\mid \bfA)$ and applying Jensen's inequality to each
term yields
\[
d(\bfB_n,\bfB_0)
\le
\int d(\bfB_n,\bfB')\,\Pi_{\bfB}(d\bfB'\mid \bfA)
+
\int d(\bfB',\bfB_0)\,\Pi_{\bfB}(d\bfB'\mid \bfA)
\le
\sqrt{F(\bfB_n)}+\sqrt{F(\bfB_0)}.
\]
Hence, for all sufficiently large $n$,
\[
d(\bfB_n,\bfB_0)\le \sqrt{F(\bfB_0)+1}+\sqrt{F(\bfB_0)},
\]
so the sequence is bounded in $d$.

Next, choose centered factors $\bfB_n=\bfY_n\bfY_n^\top$ and $\bfB_0=\bfY_0\bfY_0^\top$ with
$\bfY_n^\top\bfone_n=\bfY_0^\top\bfone_n=\bfzero$. By definition of the quotient distance, for each
$n$ there exists $\bfR_n\in O(r)$ such that
\[
\|\bfY_n-\bfY_0\bfR_n\|_F \le d(\bfB_n,\bfB_0)+\frac{1}{n}.
\]
Therefore,
\[
\|\bfY_n\|_F
\le
\|\bfY_n-\bfY_0\bfR_n\|_F + \|\bfY_0\bfR_n\|_F
\le
d(\bfB_n,\bfB_0)+\frac{1}{n}+\|\bfY_0\|_F.
\]
Since $\{\bfB_n\}$ is bounded in $d$, the sequence $\{\bfY_n\}$ is bounded in the finite-dimensional
Euclidean space $\{\bfY\in\bbR^{n\times r}:\bfY^\top\bfone_n=\bfzero\}$. Hence, by the
Bolzano--Weierstrass theorem, there exists a subsequence $\{\bfY_{n_k}\}$ and a limit $\bfY$ such that
\[
\bfY_{n_k}\to \bfY \qquad\text{in Frobenius norm}.
\]
Because the centering constraint is closed, $\bfY^\top\bfone_n=\bfzero$. Define
\[
\BF:=\bfY\bfY^\top\in\calMnr.
\]
Then, by taking the competitor $\bfR=I_r$,
\[
d(\bfB_{n_k},\BF)
\le
\|\bfY_{n_k}-\bfY\|_F \to 0.
\]

It remains to show that $\BF$ minimizes $F$. Since $\sup_k d(\bfB_{n_k},\bfB_0)<\infty$, there
exists a constant $C<\infty$ such that
\[
d(\bfB_{n_k},\bfB')\le d(\bfB_{n_k},\bfB_0)+d(\bfB_0,\bfB')\le C+d(\bfB_0,\bfB')
\]
for all $k$ and all $\bfB'$. Therefore,
\[
d^2(\bfB_{n_k},\bfB')\le 2C^2+2d^2(\bfB_0,\bfB'),
\]
and the right-hand side is integrable under $\Pi_{\bfB}(\cdot\mid \bfA)$ by the finite
second-moment assumption. Since $d(\bfB_{n_k},\bfB')\to d(\BF,\bfB')$ pointwise in $\bfB'$,
dominated convergence gives
\[
F(\bfB_{n_k}) \to F(\BF).
\]
Because $\{\bfB_{n_k}\}$ is a minimizing subsequence, it follows that
\[
F(\BF)=\inf_{\bfB\in\calMnr} F(\bfB).
\]
Thus $\BF$ is a \Frechet mean of $\Pi_{\bfB}(\cdot\mid \bfA)$.
\qed

\subsubsection*{Proof of Theorem 4}
Let $\gamma$ be an optimal coupling of $P$ and $Q$, so that
\[
\int d^2(\bfB_1,\bfB_2)\,\gamma(d\bfB_1,d\bfB_2)=W_2^2(P,Q)=w^2.
\]
Fix any $\bfB\in\calMnr$. Using the identity $|a^2-b^2|=|a-b|(a+b)$ and the triangle inequality,
we obtain
\begin{align*}
	\big|d^2(\bfB,\bfB_1)-d^2(\bfB,\bfB_2)\big|
	&=
	\big|d(\bfB,\bfB_1)-d(\bfB,\bfB_2)\big|
	\big(d(\bfB,\bfB_1)+d(\bfB,\bfB_2)\big) \\
	&\le
	d(\bfB_1,\bfB_2)\big(d(\bfB,\bfB_1)+d(\bfB,\bfB_2)\big).
\end{align*}
Integrating with respect to $\gamma$ and applying Cauchy--Schwarz gives
\begin{align*}
	|F_P(\bfB)-F_Q(\bfB)|
	&=
	\left|
	\int d^2(\bfB,\bfB_1)\,\gamma(d\bfB_1,d\bfB_2)
	-
	\int d^2(\bfB,\bfB_2)\,\gamma(d\bfB_1,d\bfB_2)
	\right| \\
	&\le
	\int d(\bfB_1,\bfB_2)\big(d(\bfB,\bfB_1)+d(\bfB,\bfB_2)\big)\,\gamma(d\bfB_1,d\bfB_2) \\
	&\le
	\left(\int d^2(\bfB_1,\bfB_2)\,d\gamma\right)^{1/2}
	\left(\int \big(d(\bfB,\bfB_1)+d(\bfB,\bfB_2)\big)^2\,d\gamma\right)^{1/2} \\
	&\le
	w\,\sqrt{2F_P(\bfB)+2F_Q(\bfB)}.
\end{align*}
Thus, for every $\bfB\in\calMnr$,
\begin{equation}\label{eq:FPFQ-bound}
	|F_P(\bfB)-F_Q(\bfB)|\le w\,\sqrt{2F_P(\bfB)+2F_Q(\bfB)}.
\end{equation}

Now let
\[
\delta := d(\mu_P,\mu_Q),\qquad
M := \sqrt{F_P(\mu_P)}+\sqrt{F_Q(\mu_Q)}.
\]
By the quadratic growth condition,
\begin{equation}\label{eq:qgrowth-muQ}
	\frac{\lambda}{2}\delta^2
	\le
	F_P(\mu_Q)-F_P(\mu_P).
\end{equation}
Since $\mu_Q$ minimizes $F_Q$, we have $F_Q(\mu_Q)\le F_Q(\mu_P)$, and therefore
\[
F_P(\mu_Q)-F_P(\mu_P)
\le
|F_P(\mu_Q)-F_Q(\mu_Q)| + |F_P(\mu_P)-F_Q(\mu_P)|.
\]
Applying \eqref{eq:FPFQ-bound} at $\bfB=\mu_Q$ and $\bfB=\mu_P$ yields
\begin{equation}\label{eq:two-term-bound}
	\frac{\lambda}{2}\delta^2
	\le
	w\sqrt{2F_P(\mu_Q)+2F_Q(\mu_Q)}
	+
	w\sqrt{2F_P(\mu_P)+2F_Q(\mu_P)}.
\end{equation}

We next bound the two square-root terms. By the triangle inequality,
\[
d(\mu_Q,\bfB)\le d(\mu_P,\bfB)+\delta
\qquad\text{for all }\bfB\in\calMnr.
\]
Applying Minkowski's inequality under $P$ gives
\[
\sqrt{F_P(\mu_Q)}
=
\left(\int d^2(\mu_Q,\bfB)\,P(d\bfB)\right)^{1/2}
\le
\left(\int d^2(\mu_P,\bfB)\,P(d\bfB)\right)^{1/2}+\delta
=
\sqrt{F_P(\mu_P)}+\delta.
\]
Similarly,
\[
\sqrt{F_Q(\mu_P)}\le \sqrt{F_Q(\mu_Q)}+\delta.
\]
Therefore,
\begin{align*}
	\sqrt{2F_P(\mu_Q)+2F_Q(\mu_Q)}
	&\le
	\sqrt{2}\big(\sqrt{F_P(\mu_Q)}+\sqrt{F_Q(\mu_Q)}\big)
	\le
	\sqrt{2}(M+\delta),\\
	\sqrt{2F_P(\mu_P)+2F_Q(\mu_P)}
	&\le
	\sqrt{2}\big(\sqrt{F_P(\mu_P)}+\sqrt{F_Q(\mu_P)}\big)
	\le
	\sqrt{2}(M+\delta).
\end{align*}
Substituting these bounds into \eqref{eq:two-term-bound} gives
\[
\frac{\lambda}{2}\delta^2 \le 2\sqrt{2}\,w(M+\delta).
\]
Rearranging,
\[
\delta^2 - a\delta - aM \le 0,
\qquad
a:=\frac{4\sqrt{2}}{\lambda}w.
\]
Since $\delta\ge 0$, this implies
\[
\delta \le \frac{a+\sqrt{a^2+4aM}}{2}.
\]
Using $\sqrt{a^2+4aM}\le a+2\sqrt{aM}$ for $a,M\ge 0$, we obtain
\[
\delta \le a+\sqrt{aM}.
\]
Substituting the definition of $a$ and $M$ yields
\[
d(\mu_P,\mu_Q)
\le
\frac{4\sqrt{2}}{\lambda}\,w
+
2^{5/4}\sqrt{\frac{w}{\lambda}}\,
\Big(\sqrt{F_P(\mu_P)}+\sqrt{F_Q(\mu_Q)}\Big)^{1/2}.
\]
This proves the stated bound. The final claim follows immediately, since the right-hand side tends
to zero as $w=W_2(P,Q)\to 0$.
\qed

\subsubsection*{Proof of Theorem 5}
Let
\[
F(\bfB):=\int d^2(\bfB,\bfB')\,\Pi_{\bfB}(d\bfB'\mid \bfA)
\]
denote the population \Frechet functional associated with the induced posterior, and let
$\BF$ be any \Frechet mean of $\Pi_{\bfB}(\cdot\mid \bfA)$. By the minimizing property of
$\BF$,
\[
F(\BF)\le F(\bfB^\star)
=
\int d^2(\bfB^\star,\bfB)\,\Pi_{\bfB}(d\bfB\mid \bfA).
\]
By the posterior mean-square concentration assumption in the manuscript, the right-hand side
converges to zero in probability. Hence
\[
F(\BF)\xrightarrow{P}0.
\]

To control $d(\BF,\bfB^\star)$, fix $\bfB$ and apply the triangle inequality:
\[
d(\BF,\bfB^\star)\le d(\BF,\bfB)+d(\bfB,\bfB^\star).
\]
Squaring both sides and using $(a+b)^2\le 2a^2+2b^2$ yields
\[
d^2(\BF,\bfB^\star)\le 2d^2(\BF,\bfB)+2d^2(\bfB,\bfB^\star).
\]
Integrating with respect to $\Pi_{\bfB}(d\bfB\mid \bfA)$ gives
\[
d^2(\BF,\bfB^\star)
\le
2F(\BF) + 2F(\bfB^\star)
\le
4F(\bfB^\star).
\]
Since $F(\bfB^\star)\xrightarrow{P}0$, it follows that
\[
d(\BF,\bfB^\star)\xrightarrow{P}0.
\]

It remains to transfer this to pairwise distances. Choose centered factors
\[
\BF=\bfY_F\bfY_F^\top,\qquad \bfB^\star=\bfY^\star{\bfY^\star}^\top,
\]
with
\[
\bfY_F^\top\bfone_n=(\bfY^\star)^\top\bfone_n=\bfzero,
\]
and align them so that
\[
\|\bfY_F-\bfY^\star\|_F = d(\BF,\bfB^\star).
\]
Let $\bfy_{F,i}$ and $\bfy_i^\star$ denote their $i$th rows. Then
\begin{align*}
	|D_{ij}(\BF)-D_{ij}(\bfB^\star)|
	&=
	\big|\|\bfy_{F,i}-\bfy_{F,j}\|-\|\bfy_i^\star-\bfy_j^\star\|\big| \\
	&\le
	\|(\bfy_{F,i}-\bfy_{F,j})-(\bfy_i^\star-\bfy_j^\star)\| \\
	&\le
	\|\bfy_{F,i}-\bfy_i^\star\|+\|\bfy_{F,j}-\bfy_j^\star\| \\
	&\le
	2\|\bfY_F-\bfY^\star\|_F \\
	&=
	2d(\BF,\bfB^\star).
\end{align*}
Since $d(\BF,\bfB^\star)\xrightarrow{P}0$, we conclude that
\[
D_{ij}(\BF) \xrightarrow{P} D_{ij}(\bfB^\star).
\]
\qed

\subsubsection*{Proof of Corollary 6}
By Theorem~5 and the assumption on $\bar{\alpha}$, we have
\[
D_{ij}(\BF) \xrightarrow{P} D_{ij}(\bfB^\star)
\quad\text{and}\quad
\bar{\alpha}\xrightarrow{P}\alpha^\star.
\]
Therefore,
\[
\bar{\alpha}-D_{ij}(\BF)
\xrightarrow{P}
\alpha^\star-D_{ij}(\bfB^\star).
\]
Since $g$ is continuous, the continuous mapping theorem yields
\[
g\!\left(\bar{\alpha}-D_{ij}(\BF)\right)
\xrightarrow{P}
g\!\left(\alpha^\star-D_{ij}(\bfB^\star)\right).
\]
\qed

\subsubsection*{Proof  of Theorem 7}
Fix a node $i$ and another node $j\neq i$. Let $\bfB$ denote a random matrix drawn from
$\Pi_{\bfB}(\cdot\mid \bfA)$. Since $D_{ij}(\bfB^\star)$ is a constant,
\[
\mathrm{Var}\{D_{ij}(\bfB)\}
\le
\bbE\big[\big(D_{ij}(\bfB)-D_{ij}(\bfB^\star)\big)^2\,\big|\,\bfA\big].
\]
As in the proof of Theorem~5, the pairwise distance functional is Lipschitz with respect to the
quotient distance:
\[
|D_{ij}(\bfB)-D_{ij}(\bfB^\star)|\le 2d(\bfB,\bfB^\star).
\]
Therefore,
\[
\mathrm{Var}\{D_{ij}(\bfB)\}
\le
4\,\bbE\big[d^2(\bfB,\bfB^\star)\,\big|\,\bfA\big]
=
4\int d^2(\bfB,\bfB^\star)\,\Pi_{\bfB}(d\bfB\mid \bfA).
\]
By the posterior mean-square concentration assumption in the manuscript, the right-hand side
converges to zero in probability. Hence
\[
\mathrm{Var}\{D_{ij}(\bfB)\}\xrightarrow{P}0
\qquad\text{for every } j\neq i.
\]

Finally, recalling the definition
\[
U_i = \frac{1}{n-1}\sum_{j\neq i}\mathrm{Var}\{D_{ij}(\bfB)\},
\]
and noting that the number of terms in the sum is finite, we conclude that
\[
U_i\xrightarrow{P}0.
\]
\qed

\newpage
\subsection*{Additional Methodological Details}

\subsubsection*{Algorithm for Computing the Sample \Frechet Mean}

\begin{algorithm}[ht!]
	\caption{Computation of the sample \Frechet mean $\BhatF$ from posterior draws}
	\label{alg:frechet}
	\begin{algorithmic}[1]
		\REQUIRE Centered factors $\bfY^{(1)},\ldots,\bfY^{(M)}\in\bbR^{n\times r}$ with $(\bfY^{(m)})^\top\bfone_n=\bfzero$, tolerance $\epsilon>0$
		\ENSURE Factor $\hat{\bfY}$ and sample \Frechet mean $\BhatF=\hat{\bfY}\hat{\bfY}^\top$
		\STATE Initialize $\bfY$ (e.g., from the rank-$r$ eigendecomposition of $(1/M)\sum_{m=1}^M \bfB^{(m)}$)
		\REPEAT
		\FOR{$m=1,\ldots,M$}
		\STATE Compute the SVD $(\bfY^{(m)})^\top\bfY=\bfU_m\bfSigma_m\bfV_m^\top$
		\STATE Compute $\bfR_m^\star=\bfU_m\bfV_m^\top$ and $\tilde{\bfY}^{(m)}=\bfY^{(m)}\bfR_m^\star$
		\ENDFOR
		\STATE Set $\bar{\bfY}\leftarrow \tfrac{1}{M}\sum_{m=1}^M \tilde{\bfY}^{(m)}\quad\text{and}\quad \bfZ\leftarrow \bar{\bfY}-\bfY$
		\STATE Compute $\bfZ_{\mathrm{hor}}\leftarrow \mathrm{Proj}^{\mathrm{hor}}_{\bfY}(\bfZ)$
		\STATE Choose a step size $\eta>0$ and update
		\[
		\bfY\leftarrow \mathrm{Retr}_{\bfY}(\eta\,\bfZ_{\mathrm{hor}})
		\]
		\UNTIL{$\|\eta\,\bfZ_{\mathrm{hor}}\|_F<\epsilon$}
		\STATE Set $\hat{\bfY}\leftarrow \bfY$
		\STATE \textbf{return} $\hat{\bfY}$ and $\BhatF=\hat{\bfY}\hat{\bfY}^\top$
	\end{algorithmic}
\end{algorithm}

\subsubsection*{Derivation of the Classical MDS Identity}

A brief derivation is as follows. Write $\tilde \bfX=\bfH\bfX$, so
$\bfB=\tilde \bfX\tilde \bfX^\top$. For any $i,j \in [n]\times [n]$,
\[
\Delta_{ij}=\|\tilde \bfx_i-\tilde \bfx_j\|^2
=\tilde \bfx_i^\top\tilde \bfx_i+\tilde \bfx_j^\top\tilde \bfx_j-2\tilde \bfx_i^\top\tilde \bfx_j.
\]
In matrix form,
\[
\Delta=\mathrm{diag}(\bfB)\bfone_n^\top+\bfone_n\mathrm{diag}(\bfB)^\top-2\bfB.
\]
Premultiplying and postmultiplying by $\bfH$ removes the two terms involving $\bfone_n$, yielding
\[
\bfH\Delta \bfH = -2\bfH\bfB \bfH = -2\bfB,
\]
because $\bfH\bfB=\bfB$ and $\bfB\bfH=\bfB$.

\newpage
\subsection*{Additional Empirical Details}

\subsubsection*{Additional Simulation Summaries}

Table~\ref{tab:sim-summary} reports numerical summaries corresponding to the simulation
diagnostics in the main text. Specifically, it summarizes the reference-sensitivity index
$S_{\mathrm{ref}}$ by regime together with group-specific summaries of the node-level uncertainty
index $U_i$ and the invariant node-wise loss $L_i$ for the representative weak-regime dataset.
These numerical results support the same qualitative conclusions as the main-text figures:
reference sensitivity is substantially larger in the weakly identified regime, the bridge group has
the largest uncertainty, and $U_i$ is positively associated with $L_i$.

\begin{landscape}
	\begin{table}[p]
		\centering
		\small
		\begin{tabular}{lccccccc}
			\toprule
			& \multicolumn{2}{c}{\textbf{Global regime summary}} & \multicolumn{3}{c}{\textbf{$U_i$ by group}} & \multicolumn{2}{c}{\textbf{$L_i$ by group}} \\
			\cmidrule(lr){2-3}\cmidrule(lr){4-6}\cmidrule(lr){7-8}
			\textbf{Regime} & \textbf{$S_{\mathrm{ref}}$} & \textbf{Corr($U_i,L_i$)} & \textbf{Left} & \textbf{Bridge} & \textbf{Right} & \textbf{Left} & \textbf{Bridge/Right} \\
			\midrule
			Well-identified
			& 4.633 (8.173)
			& --
			& --
			& --
			& --
			& --
			& -- \\
			Weakly identified
			& 9.782 (5.811)
			& 0.712
			& 0.436 (0.090)
			& 0.532 (0.152)
			& 0.436 (0.101)
			& 0.899 (0.385)
			& 1.181 (0.644) / 0.806 (0.325) \\
			\bottomrule
		\end{tabular}
		\caption{Quantitative summaries of the simulation study. The reference-sensitivity index
			$S_{\mathrm{ref}}$ is reported as mean (sd) across simulated networks within each regime. For the
			representative weak-regime dataset, group-specific values report mean (sd) of the node-level
			uncertainty index $U_i$ and invariant node-wise loss $L_i$. The correlation column reports the
			Pearson correlation between $U_i$ and $L_i$.}
		\label{tab:sim-summary}
	\end{table}
\end{landscape}

\subsubsection*{Coauthorship Preprocessing Details}

The coauthorship analysis begins from the MATLAB file \texttt{CoauAdjFinal.mat}, which stores an
adjacency matrix $A$ together with the corresponding author names. This file is produced from the
author--paper incidence data in the MADStat repository available at \url{https://github.com/ZhengTracyKe/MADStat} by first forming the weighted coauthorship matrix
$A^{\mathrm{wt}} = XX^\top$, where $X$ is the author--paper incidence matrix, then thresholding the
weights so that
\[
A_{ij}=1 \quad\Longleftrightarrow\quad \text{authors } i \text{ and } j \text{ coauthored at least three papers},
\]
and setting $A_{ii}=0$. The resulting thresholded graph is then restricted to its largest connected
component before being saved in \texttt{CoauAdjFinal.mat}. In our case, this network contains
4{,}383 authors.

Community labels are taken from the companion file
\texttt{CommunityResults\_firstlayer.mat}, which stores the first-layer partition used in
\citet{ji_2022_CocitationCoauthorshipNetworks}. These labels are obtained by applying SCORE to the
adjacency matrix with diagonal regularization and then relabeling the six detected communities to
match the interpretation in that paper. In the R implementation, the author ordering in
\texttt{CoauAdjFinal.mat} and \texttt{CommunityResults\_firstlayer.mat} was checked using the
stored \texttt{authorNames} fields, and all subsequent preprocessing was carried out only after
verifying that the node orderings agreed.

We then restricted attention to the three first-layer biostatistics communities with labels
2, 4, and 6, corresponding to Biostatistics (Europe), Biostatistics (UNC), and Biostatistics
(Michigan), respectively. Their sizes are 202, 673, and 264, yielding an induced subgraph on
1{,}139 authors. This induced subgraph is disconnected: it contains 64 connected components, of
which the largest has 980 nodes. All subsequent model fitting and posterior analysis are based on
this 980-node component.

For computational feasibility, the final fitted network was obtained by trimming the 980-node
component to a connected 400-node subgraph. In the analysis code, nodes were ranked by degree
within the 980-node component, the highest-degree 400 nodes were retained, and the largest
connected component of this retained subgraph was used for model fitting. The latent space model and
all quotient-based summaries reported in the main text are therefore based on this final connected
400-node network.

\begin{table}[h]
	\centering
	\small
	\begin{tabular}{lc}
		\toprule
		Quantity & Size \\
		\midrule
		Biostatistics (Europe) & 202 \\
		Biostatistics (UNC) & 673 \\
		Biostatistics (Michigan) & 264 \\
		Selected induced subgraph & 1{,}139 \\
		Largest connected component & 980 \\
		Final fitted network & 400 \\
		\bottomrule
	\end{tabular}
	\caption{Sizes of the selected coauthorship subnetwork at each preprocessing stage.}
	\label{tab:coauthorship-preprocess}
\end{table}

\end{document}